\documentclass[a4paper,fleqn,usenatbib]{mnras}
\usepackage[T1]{fontenc}
\usepackage{ae,aecompl}
\usepackage{amsmath,amssymb}

\usepackage{color, soul}
\usepackage{hyperref}
\usepackage{xcolor}
\usepackage{graphics,graphicx}
\usepackage{epsf} 
\usepackage{longtable}
\usepackage{deluxetable}
\usepackage{ulem}
\def\arcsec{$^{\prime\prime}$}
\def\~{$\sim$}

\def\arcsec{$^{\prime\prime}$}

\definecolor{darkgreen}{rgb}{0.1, 0.2, 0.13}
\title[H{\sc i} content of  mid-infrared bright BCDs]{H{\sc i} content of selected mid-infrared bright, starburst blue compact dwarf galaxies}
\author[Chandola Y., et al.]{Yogesh Chandola$^{1,2,3}$\thanks{E-mail: yogesh.chandola@pmo.ac.cn}, Di Li$^{1,4,5}$\thanks{E-mail: dili@nao.cas.cn}, 
	Chao-Wei Tsai$^{1,6,7}$\thanks{E-mail: cwtsai@nao.cas.cn}, 
	Guodong Li$^{1,7}$, 
	Yingjie Peng$^{8,9}$, \newauthor
	Pei Zuo$^{1,9,10}$, Travis McIntyre$^{1}$,  Yin-Zhe Ma$^{11,2}$\thanks{E-mail: mayinzhe@sun.ac.za},
	Daniel Stern$^{12}$, 
	Roger Griffith$^{12}$, \newauthor
    Thomas Jarrett$^{13}$,
	Peter Eisenhardt$^{12}$, Chantal Balkowski$^{14}$
	\vspace{2mm}
	\\
	$^{1}$CAS Key Laboratory of FAST, National Astronomical Observatories, Chinese Academy of Sciences, Beijing-100101, China\\
	$^{2}$Purple Mountain Observatory, Chinese Academy of Sciences (CAS), 10, Yuan Hua Road, Qixia District, Nanjing, 210023, China \\
	$^{3}$Inter-University Centre for Astronomy and Astrophysics (IUCAA), Post bag 4, Ganeshkhind, Pune, 411007, India \\
	$^{4}$Research Center for Intelligent Computing Platforms, Zhejiang Laboratory, Hangzhou 311100, China\\
	$^{5}$NAOC-UKZN Computational Astrophysics Centre (NUCAC), University of KwaZulu-Natal, Durban, 4000, South Africa \\
	$^{6}$Institute for Frontiers in Astronomy and Astrophysics, Beijing Normal University,  Beijing 102206, China\\
	$^{7}$University of Chinese Academy of Sciences, Beijing 100049, People's Republic of China\\
	$^{8}$Department of Astronomy, School of Physics, Peking University, 5 Yiheyuan Road, Beijing 100871, People's Republic of China\\
	$^{9}$Kavli Institute for Astronomy and Astrophysics, Peking University, 5 Yiheyuan Road, Beijing 100871, People's Republic of China\\
	$^{10}$International Centre for Radio Astronomy Research (ICRAR), University of Western Australia, 35 Stirling Highway, Crawley,\\ WA 6009, Australia\\
        $^{11}$Department of Physics, Stellenbosch University, Matieland 7602, South Africa \\
	$^{12}$Jet Propulsion Laboratory, California Institute of Technology, 4800 Oak Grove Drive, M/S 169-327, Pasadena, CA 91109, USA\\
	$^{13}$Department of Astronomy, University of Cape Town, Private Bag X3, Rondebosch, 7701, South Africa\\
	$^{14}$Observatorie de Paris, 61 Av. de l'Observatoire, 75014 Paris, France}
\date{\today}
\pubyear{2023}

\begin{document}
\label{firstpage}
\maketitle

\begin{abstract}
We report measurements of H{\sc i} content in 11 nearby, actively star-forming, blue compact dwarf galaxies (BCDs) from 21 cm observations with the  Arecibo telescope. These BCDs, selected by their red (W2[4.6 $\mu$m]$-$W3[12 $\mu$m]$>$3.8 mag) and bright mid-infrared (MIR) emission (W4[22 $\mu$m]$<$ 7.6 mag), have high specific star formation rates (median sSFR $\sim$10$^{-7.8}$ yr$^{-1}$), similar to high redshift galaxies. H{\sc i} emission was detected in six sources. We analyze our new detections in the context of previous H{\sc i} observations of 218 dwarf irregulars (dIs) and BCDs in the literature. The $M_{\rm HI}$-$M_{\ast}$ relation resulting from our observations confirms the dominating fraction of H{\sc i} gas among baryons in galaxies with lower stellar masses.  This Arecibo BCD sample has significantly lower median H{\sc i}  depletion timescales ($\tau_{\rm HI}\sim$ 0.3 Gyr) than  other dIs/BCDs ($\sim$ 6.3 Gyr) in the literature. The majority of the sources (10/11) in the Arecibo sample are very red in W1[3.4 $\mu$m]$-$W2[4.6 $\mu$m] colour ($>$ 0.8 mag) implying the presence of warm dust. We investigate the relation of $\tau_{\rm HI}$ with stellar mass ($M_{\ast}$) and sSFR. We find that $\tau_{\rm HI}$ is significantly anti-correlated with $M_{\ast}$ for higher sSFR ($>$10$^{-8.5}$ yr$^{-1}$) and with sSFR for higher stellar mass ($>10^{7.5}\,{\rm M}_{\odot}$) dwarf galaxies. The high sSFR for the BCDs in the Arecibo observed sample is mainly due to their high atomic gas star formation efficiency (SFE) or low $\tau_{\rm HI}$. The low $\tau_{\rm HI}$ or high SFE in these sources is  possibly due to runaway star formation in compact and dense super star clusters. 
\end{abstract}
\section{Introduction}
\label{sec1}
Blue Compact Dwarf Galaxies (BCDs) are galaxies characterized spectroscopically by intense emission lines and blue, fainter optical continuum ($M_{\rm B}$ $>$ $-$18; \citealt{1966ApJ...143..192Z}; \citealt{2003ApJS..147...29G}). Their metallicities are low, ($Z$ $<$ 1/2 $Z_{\odot}$; \citealt{1999AJ....117.2789H}), and they have dramatically different physical properties compared to normal dwarf galaxies (\citealt{1966ApJ...143..192Z}; \citealt{2003ApJS..147...29G}). The great majority of BCDs are metal-poor, although BCDs with extremely low metallicity ($<$ 1/10 $Z_{\odot}$) are rare \citep{2012A&A...546A.122I}. There have been recent efforts to increase the numbers of extremely low metallicity BCDs \citep{2003ApJ...593L..73K, 2011ApJ...743...77M, 2012A&A...546A.122I, 2015MNRAS.448.2687J, 2016ApJ...822..108H, 2016ApJ...819..110S,  2017ApJ...835..159S, 2017A&A...599A..65G, 2018MNRAS.473.1956I, 2018ApJ...863..134H}. Generally, all BCDs host recent star formation, their optical blue colour indicating  a starburst no older than a few Myr \citep{1999Ap&SS.265..489K}. Some BCDs have stars forming in compact regions ($<$ 50 pc) while others form stars in spatially extended regions ($>$ 100 pc) \citep{2004A&A...421..555H}. This dichotomy of star formation activity in BCDs can affect the temperature of surrounding dust, and thus their mid-infrared (MIR) photometry from the \textit{Wide-Infrared Survey Explorer} (\textit{WISE}; \citealt{2010AJ....140.1868W}). The stronger MIR emission in compact active star-forming BCDs is dominated by hot (200-1500 K) small dust grains (size $<$ 100 \AA) heated by the starburst \citep{2001AJ....122.1736D, 2006ApJ...639..157W}. Because of hot dust emission at $\sim$10 $\mu$m, these BCDs have redder colours across \textit{WISE} 3.4--12 $\mu$m bands (W1-W3: 3.4, 4.6, 12 $\mu$m). This provides a way to select  actively star-forming, low metallicity ($\lesssim$ 1/8 $Z_{\odot}$) and high specific star formation rate (sSFR $\gtrsim$10$^{-8}$ yr$^{-1}$) BCDs \citep{2011ApJ...736L..22G, 2016ApJ...832..119H, 2018ApJ...858...38S}. Objects with sSFR higher than 10$^{-8}$ yr$^{-1}$ are rare locally ( $z <$ 0.05), although similar objects such as green pea galaxies and Lyman-$\alpha$ Emitters are  commonly found at higher redshifts ($z >$ 0.1)  \citep{gawiser2007ApJ...671..278G,amorin2010ApJ...715L.128A, 2016ApJ...832..119H, 2017ApJ...847...38Y}. 

While molecular gas is directly related to star-formation activity, cold atomic H{\sc i} gas has a vital role as a fuel reservoir \citep{2012ARA&A..50..531K}. Studying atomic gas content and gas depletion timescales  provides insights into the evolution processes of galaxies. The dependence of gas to stellar mass ratios and star-formation efficiencies/depletion timescales on stellar masses and specific star formation rates  have been  investigated in recent H{\sc i} and molecular gas surveys for local massive galaxies \citep[$z <$ 0.05, $M_{\star} >$ 10$^{9}\,{\rm M}_{\odot}$;][]{2010MNRAS.408..919S, 2011MNRAS.415...61S,  2014MNRAS.443.1329H,2015ApJ...808...66J, 2017ApJS..233...22S,2018MNRAS.476..875C}. At lower stellar masses ($< 10^{9}\,{\rm M}_{\odot}$),   H{\sc i} 21cm emission from dwarf galaxies, including some
BCDs have been studied using single-dish and interferometric facilities \citep{1981ApJ...247..823T, 2004AJ....128..617T,2005A&A...434..887H,  2007A&A...462..919H, 2007A&A...464..859P, 2013A&A...558A..18F, 2016MNRAS.463.4268T, 2001AJ....121.1413P, 2004AJ....128..617T,2006MNRAS.371.1849C, 2006MNRAS.372..853E, 2008MNRAS.391..881E,2009MNRAS.397..963E, 2010MNRAS.403..295E, 2012AJ....143..133H,2012ApJ...756..113H, 2014MNRAS.445.1694L,2018A&A...612A..26A}. Most of these sources have sSFR lower than 10$^{-8}$ yr$^{-1}$ and redshifts below 0.02. In this paper, we present deep H{\sc i} observations of 11 local ($z < 0.033$, median redshift $\sim$ 0.026) MIR-bright BCDs with an ongoing major episode of star formation with most of these having sSFR $\gtrsim$ 10$^{-8}$ yr$^{-1}$. In addition to these sources, we collect information on several dwarf irregulars(dIs)/BCDs studied in H{\sc i} from literature to  provide a comprehensive study of  the relation between  H{\sc i} gas content and galaxy properties such as stellar mass and star-formation rate.

	 In this paper, we describe the sample for study in Section~\ref{sec:sample}. In Section~\ref{sec:observations}, details of observations with the Arecibo Observatory and data reduction procedure are given. Results are reported in Section~\ref{sec:results}. The methods of estimating the stellar mass and star formation rate  are included in Section~\ref{sec:smsfroh}. We discuss our results in Section~\ref{sec:analysis} and summarize them in Section~\ref{sec:summary}. Magnitudes are reported in the Vega system.  
	 In this paper, we assume a concordance cosmology with $H_0 = 70\, {\rm km}\, {\rm s}^{-1}\, {\rm Mpc}^{-1}$, $\Omega_{\rm m} = 0.3$, and $\Omega_\Lambda = 0.7$.

\section{Sample, observation and data reduction}
\label{sec:sample}
\subsection{Sample of BCDs observed with  Arecibo Observatory}
\begin{figure*}
	\center
	\includegraphics[scale=0.5]{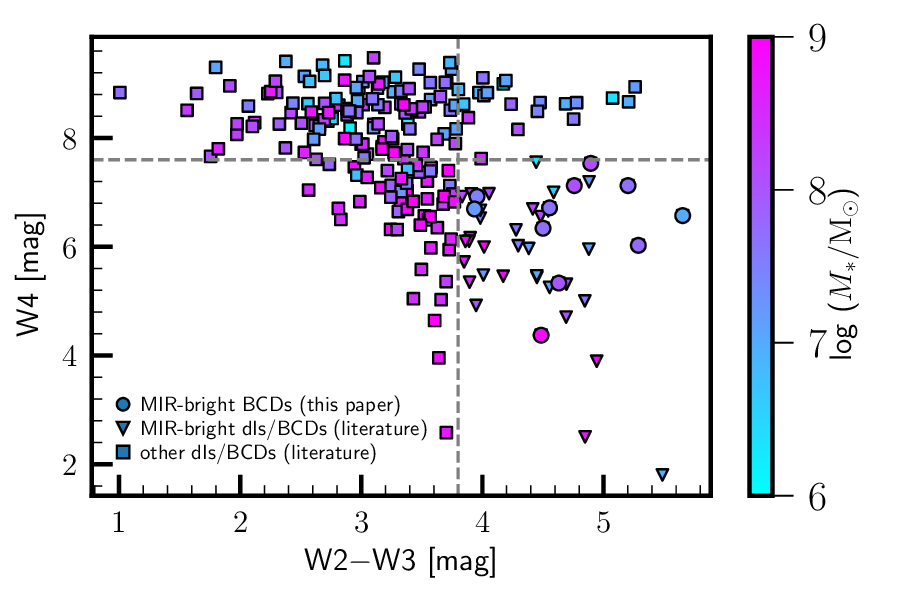} 
	\includegraphics[scale=0.5]{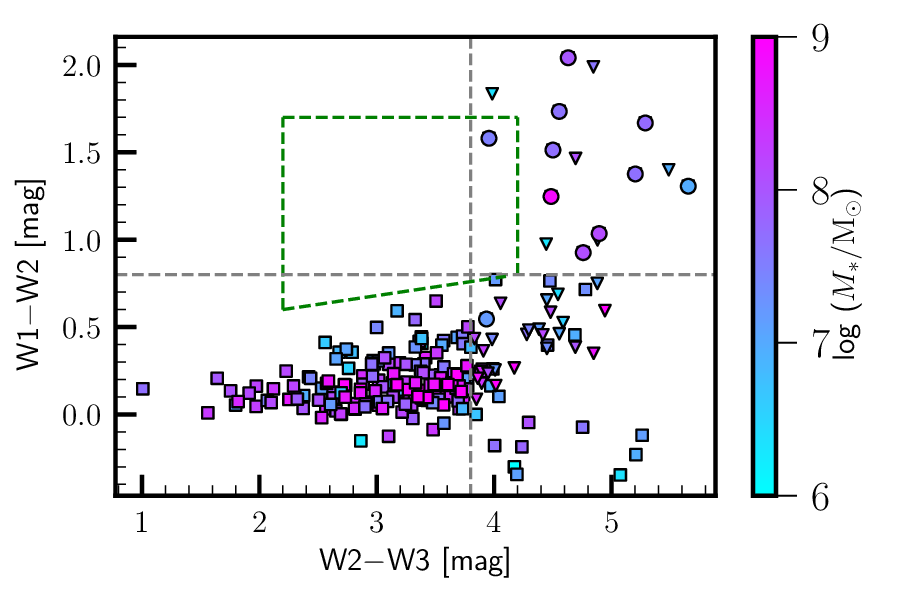}
	\caption{\textbf{Left panel:} \textit{WISE} W4 vs W2$-$W3 for 225 low-redshift starburst dwarf galaxies with data in the AllWISE catalog.  MIR-bright BCDs observed with  Arecibo  are shown with circles. MIR-bright dIs/BCDs and other dIs/BCDs from the literature are shown with downward triangles and  squares, respectively. Horizontal and vertical dashed grey lines mark our selection criteria for MIR-bright BCDs. \textbf{Right panel:} \textit{WISE} W1$-$W2 vs W2$-$W3. Symbols mean same as in the left panel. The  dashed vertical grey line represents W2$-$W3$>$3.8 mag, one of our  criteria for selecting MIR-bright BCDs. The dashed horizontal grey line represents the cutoff for selecting AGNs from \protect\cite{2012ApJ...753...30S}. The region marked with the dashed green lines represents the AGN selection criteria from \protect\cite{jarrett2011ApJ...735..112J}}. 
	\label{selectioncriteria}
\end{figure*}

\begin{figure*}
	\center
	\includegraphics[scale=0.7]{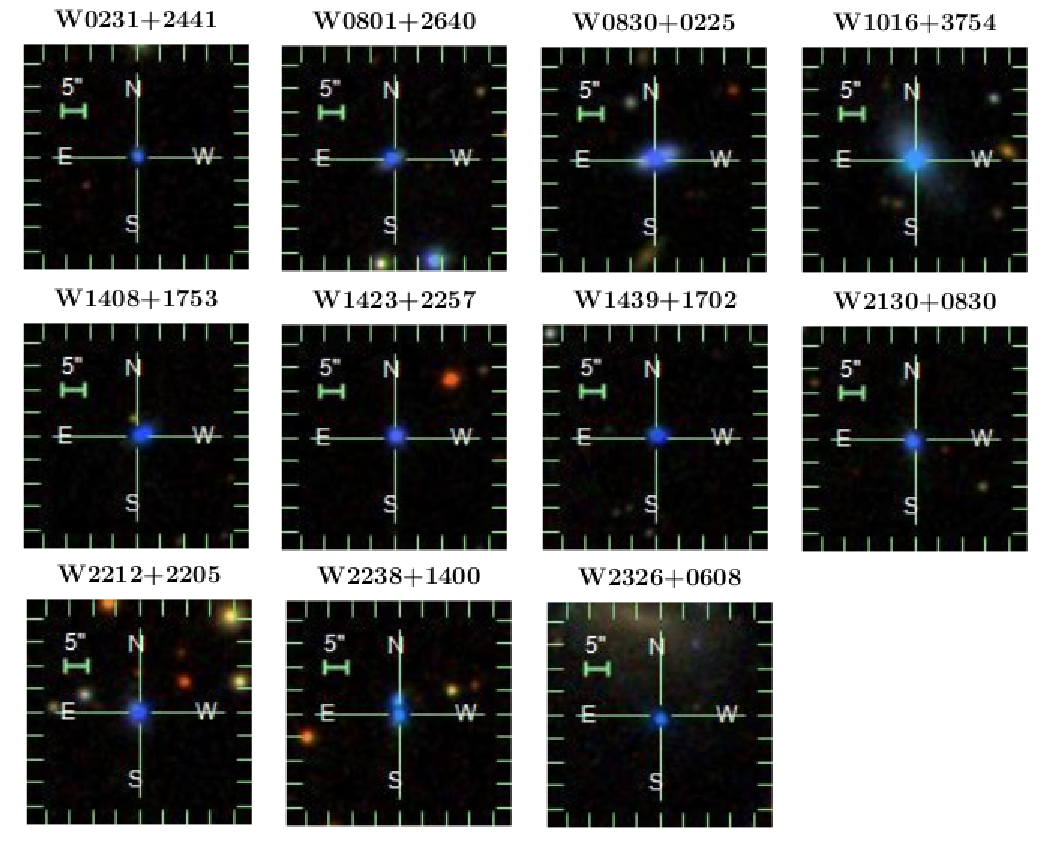} 
	\caption{Optical SDSS \textit{g, r, i} band composite images of the BCDs observed with the Arecibo. The field of view for each of these images is 50\arcsec $\times$ 50\arcsec.}
	\label{sdss}
\end{figure*}

MIR colours  can be used to select  actively star-forming, high specific star formation rate starburst BCDs.  We selected dwarf galaxies ($M_{\ast}$ $\lesssim$ 3$\times$10$^{9}\,{\rm M}_{\odot}$) with  red \textit{WISE} colour (W2$-$W3 $>$ 3.8 mag) and bright MIR emission (W4$<$ 7.6 mag equivalent to W4$>$ 7.5 mJy) from AllWISE catalog \citep{2014yCat.2328....0C}.  The red MIR colours indicate they are likely undergoing a major starburst episode \citep{2011ApJ...736L..22G, 2016ApJ...832..119H, 2018ApJ...858...38S}. The W2$-$W3 colour cut was empirically chosen to include the star-forming dwarfs discovered by the  \textit{WISE} in the early stage of the mission \citep{2012ApJ...748...80D,2012AAS...21920103T}.  While \cite{2016ApJ...832..119H} use both W1$-$W2  and W2$-$W3  colours to select BCDs, we omit using W1$-$W2 colour to select the BCDs since it is  used by some authors to select active galactic nuclei (AGN) in dwarf galaxies \citep{2018ApJ...858...38S}. Instead, we use bright W4 emission  to ensure a   high star formation rate for the sample \citep{leejongchul2013ApJ...774...62L,cluver2017ApJ...850...68C}. In Fig.~\ref{selectioncriteria}, we show the \textit{WISE} colour-colour plot and selection criteria for sources observed with  Arecibo Observatory along with  sources compiled from literature (see Section~\ref{sec:litsample} ). The Arecibo sample includes only the confirmed BCDs with their spectral energy distribution (SED) indicating no presence of an AGN. Their blue optical morphologies are shown in the optical images from Sloan Digital Sky Survey data release 12 (SDSS DR 12; \citealt{2015ApJS..219...12A}) in Fig.~\ref{sdss}. 

\subsubsection{Arecibo observations and data reduction}
\label{sec:observations}

We observed 11 nearby ($z$ $\lesssim$ 0.03) MIR-bright BCDs for their H{\sc i} emission with the Arecibo Telescope during 2012-2013. The observations  were taken during  two cycles, November-December 2012 and September-December 2013, using standard ON-OFF mode and recording in one-second integrations. Both Arecibo L-band Feed Array (ALFA) and L-band Wide receivers were used. L-band Wide data were taken with the Wide band Arecibo Pulsar Processor (WAPP) spectrometer with 4096 channels across a bandwidth of 50 MHz, giving a velocity resolution of $\sim$2.7 km s$^{-1}$. ALFA data were taken with the  spectrometer with 4096 channels across a bandwidth of 100 MHz for a velocity resolution of $\sim$ 5.4 km s$^{-1}$. The total effective integration time for each source  
are listed in Table~\ref{obslog1}.
\begin {table*}
\begin {center}
\caption {Observational details of the Arecibo search during 2012-2013 for H{\sc i} 
	emission in the BCD sample.}
\label{obslog1}
	\begin {tabular}{c c c c c c }
	\hline
	(1)              & (2)             & (3)             & (4)  & (5) &    (6)          \\ 
	Source           & R.A. & Dec. & Date            &Observed Central  & Time     \\
	&&&   of            &  Velocity        & observed   \\
	&&& observation     & [km s$^{-1}$]          & [s]       \\
	\hline                                                                         
	W0231+2441   &02:31:19.77&+24:41:23.5     & 2013 Nov. 04  &    8640       &  946              \\
	W0801+2640   &08:01:03.92&+26:40:54.4     & 2012 Nov. 08,09,11,12,13,14,15 &    7795       &   7500             \\
	W0830+0225   &08:30:44.36&+02:25:55.8     & 2013 Dec. 13 &    9450       &  1499            \\
	W1016+3754   &10:16:24.50&+37:54:45.8     & 2012 Dec. 06,07 &    1170       & 1200            \\
	W1408+1753   &14:08:16.24&+17:53:50.7    & 2013 Sep. 21, 29  &  7110   &   2700            \\
	W1423+2257   &14:23:42.85&+22:57:28.4    & 2013 Sep. 21, Sep. 29, Oct. 06 &   9900  &  2300            \\
	W1439+1702   &14:39:57.88&+17:02:17.8    & 2013 Oct. 05,06 &   9030        &  2100              \\
	W2130+0830   &21:30:05.24&+08:30:12.8     & 2013 Sep. 14,15,16  &   7800        &   2338                \\
	W2212+2205   &22:12:59.35&+22:05:05.6     & 2013 Sep. 16 Oct.19,25 Dec.08,09  &   8580        &    3900             \\
	W2238+1400   &22:38:31.09&+14:00:27.1     & 2012 Nov.06,07,08 & 6176     & 7550               \\
	W2326+0608   &23:26:03.58&+06:08:14.9     & 2013 Dec.08 &   5010        &   1100               \\
	
	\hline             
	\end {tabular}
\end{center}
\begin{center}	
Notes: Column (1): Source name. Column (2): Right Ascension. Column (3): Declination. Column (4): Date of observation. Column (5):Observed central velocity in km s$^{-1}$. Column (6): Observation time in seconds. Co-ordinates in this table are from AllWISE catalog \citep{2014yCat.2328....0C}. 
\end {center}
\end {table*}

\begin {table*}
\begin {center}
\caption {Characteristics of H{\sc i} emission profiles of observed MIR-bright BCDs}
\label{sourchar1}       
	\begin {tabular}{c c c c c c c c}
	\hline
	(1)   & (2)& (3)       & (4)          & (5)  & (6)            & (7)   & (8) \\ 
	Source & $V_{\rm opt}$&$V_\mathrm{HI}$& $W_{50}$ & $F_\mathrm{HI}$ & $S_{\rm peak}$ &$\Delta$$S_{\rm rms}$ (1$\sigma$)  & $\log M_\mathrm{HI}/{\rm M}_{\odot}$ \\
	& [km s$^{-1}$] &  [km s$^{-1}$]       &  [km s$^{-1}$]               & [mJy km s$^{-1}$]& [mJy]   &     [mJy]   &              \\
	
	\hline
	W0231+2441 & 8634  & 8671$\pm$7 & 42$\pm$14 & 50$\pm$40 &1.4 &0.7 &8.3 \\
	W0801+0264 & 7795  &  7920$\pm$5 & 96$\pm$10& 222$\pm$28 &2.4 &0.2 & 8.8 \\
	W0830+0225 & 9443  &  9413$\pm$15 & 115$\pm$30&290$\pm$70 &2.6 &0.5 & 9.1 \\
	W1016+3754 & 1169  & 1177$\pm$4 & 61$\pm$8   & 490$\pm$60 &8.1 &0.5 & 7.5 \\
	W1408+1753 & 7105  &  -      &-&-&-&0.4 &$<$ 8.2 \\  
	W1423+2257 & 9863  &  -      &-&-&-&0.5 &$<$ 8.7 \\
	W1439+1702 & 9024  &   -      &-&-&-&0.5 &$<$ 8.3 \\
	W2130+0830 & 7795  &  -      &-&-&-&0.4 &$<$ 8.1 \\
	W2212+2205 & 8574  &  -      &-&-&-&0.5 &$<$ 8.5 \\
	W2238+1400 & 6176  &  6165$\pm$16 & 93$\pm$32 & 66$\pm$23 &0.8 &0.2 & 8.1 \\
	W2326+0608&	5031  &  4996$\pm$2 & 99$\pm$4  & 1670$\pm$130 &18.7 & 0.5 & 9.3 \\
W2326+0608$^{*}$& 5031 & 5023$\pm$7 &57$\pm$15  & 50$\pm$30 &1.5 & 0.4 & 7.8 \\
	\hline                          
	\end {tabular}
	\begin{center}    
		$^{*}$Extended H{\sc i} disk, here calculated from high resolution GMRT follow-up observations  \citep{chandola2023MNRAS.523.3848C}.

		Notes:
		Column (1): Source name.
		Column (2): $V_{\rm opt}$, velocity corresponding to optical redshift (c*$z$) 
		Column (3): $V_{\rm H \sc{I}}$,  H{\sc i} velocity in km s$^{-1}$ in heliocentric frame.
		Column (4): $W_{50}$, velocity width in km s$^{-1}$ of the profile at 50\% of the peak flux level.
		Column (5): $F_{\rm HI}$, integrated flux density in mJy km s$^{-1}$. 
		Column (6): $S_{\rm peak}$, peak flux density in mJy.
		Column (7): $\Delta S_{\rm rms}$, noise level per channel in mJy.
		Column (8): log $M_{\rm HI}$, logarithm of the total H\,{\sc{i}} mass in M$_{\odot}$.   
	\end{center}
\end {center}                      
\end {table*} 
The data were reduced in IDL using both standard programs provided by the observatory and software developed for combining integrations from multiple observations with ALFA and L-band Wide receiver. Bandpasses were corrected using position switching (ON-OFF)/OFF and Hanning smoothing was applied. Baselines were automatically fit to first order, calibration was applied, and polarizations were averaged. Radio frequency interference (RFI) was identified both automatically and manually and removed from the final spectra. The most common  RFI was GPS L3 satellite interference, as many sources in the sample are located near 8500 km s$^{-1}$. L-band Wide data  were  smoothed further to align the effective velocity resolution of L-band Wide data with ALFA data at $\sim$ 10 km s$^{-1}$. Final noise levels vary from 0.2 mJy channel$^{-1}$ to 0.7 mJy channel$^{-1}$ (Table~\ref{sourchar1}), though the white noise in several spectra is dominated by baseline structure likely caused by reflections from the blocked aperture of the Arecibo telescope. The parameters, peak flux ($S_{\rm peak}$), integrated flux ($F_{\rm HI}$), H{\sc i} velocity ($V_{\rm HI}$), and Full Width at Half Maximum ($W_{50}$), listed in Table~\ref{sourchar1}, are obtained from H{\sc i} profiles using standard Arecibo software package \textit{mbmeasure}\footnote{\url{http://www.naic.edu/~rminchin/idl/mbmeasure.pro}}. The uncertainties on these parameters are calculated in the same way as in \cite{2004AJ....128...16K}.
\begin{figure*}
	\centering 
	\includegraphics[scale=0.26]{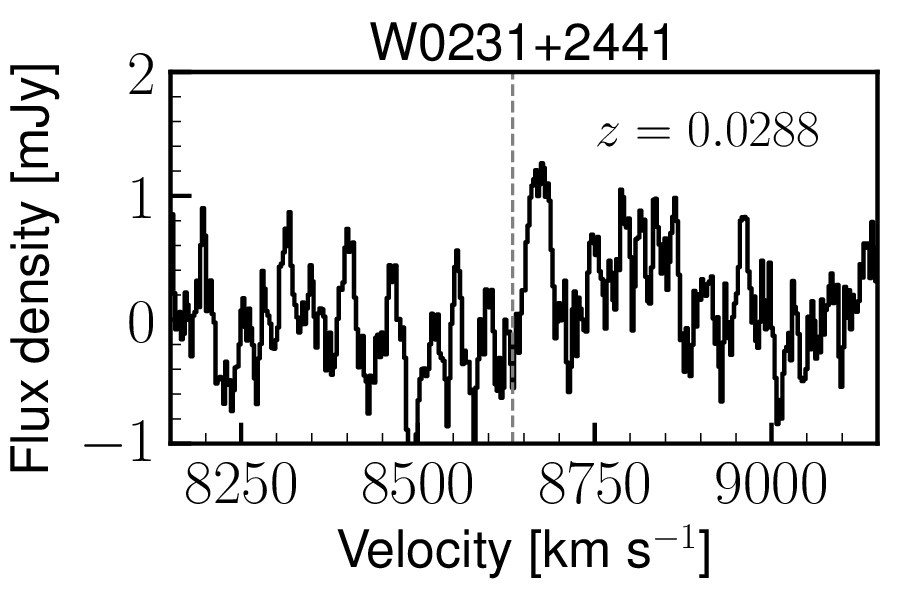}   
	\includegraphics[scale=0.26]{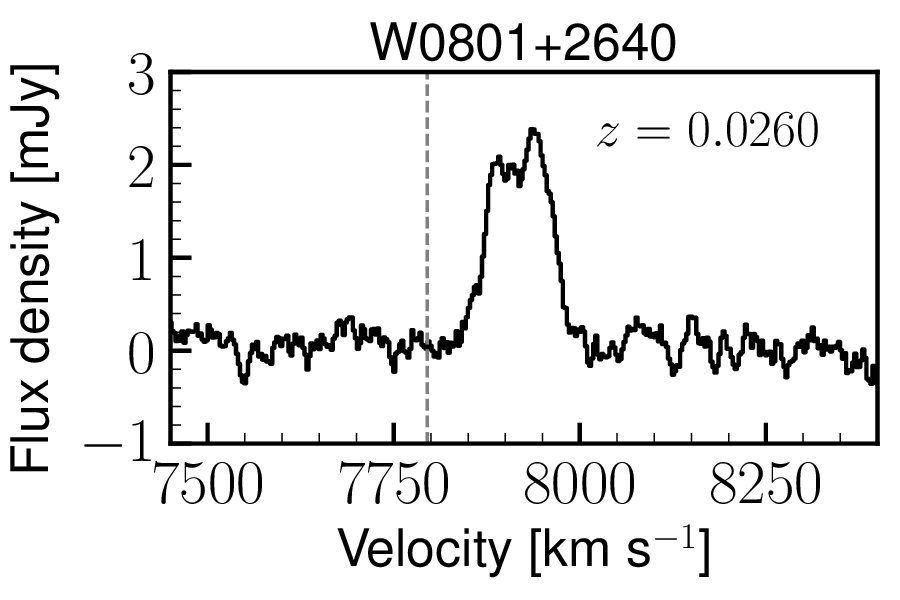}
	\includegraphics[scale=0.26]{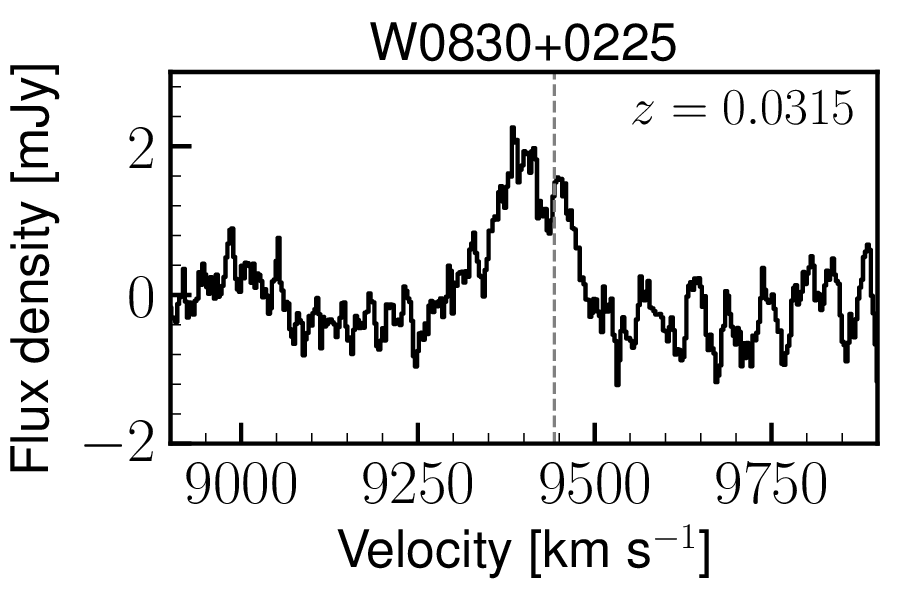}
	\includegraphics[scale=0.26]{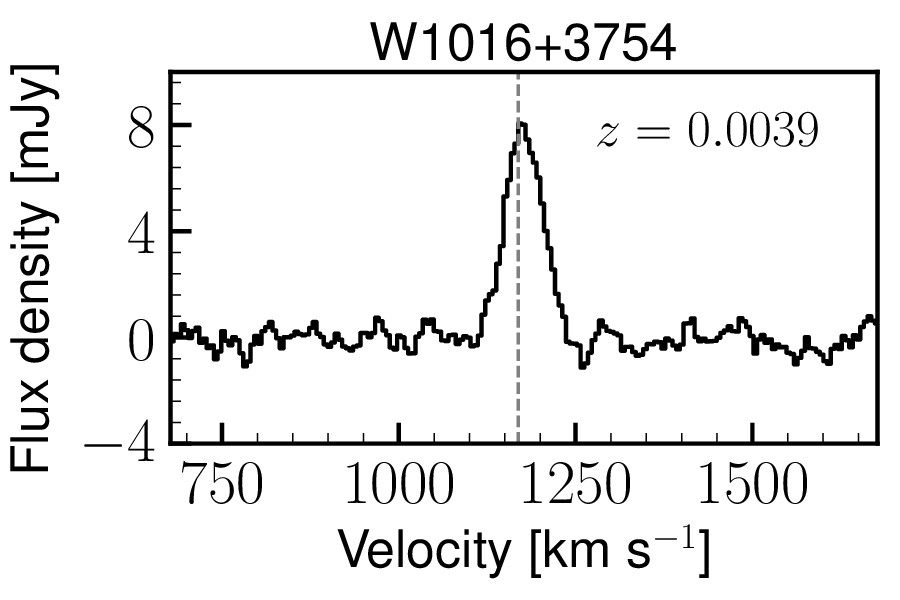}
	\includegraphics[scale=0.26]{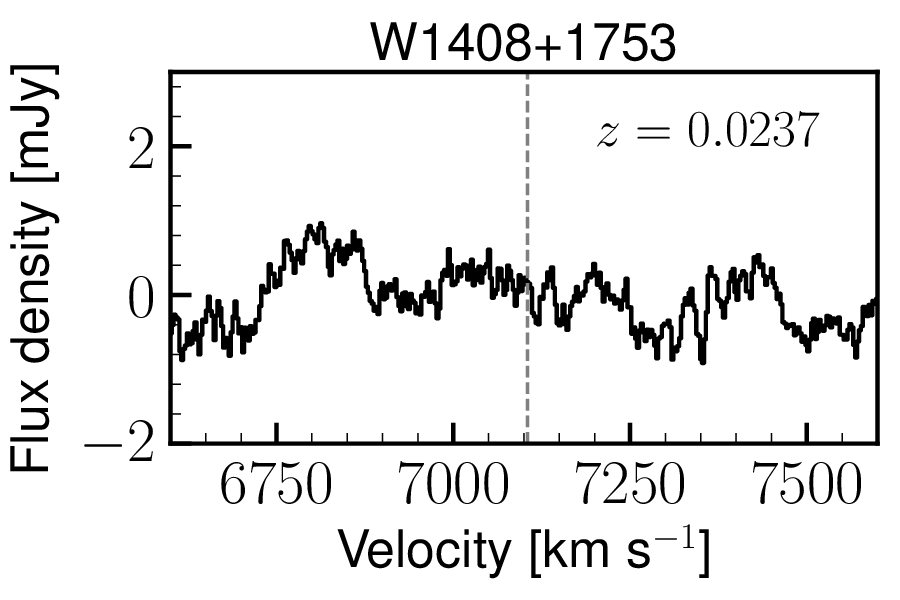}
	\includegraphics[scale=0.26]{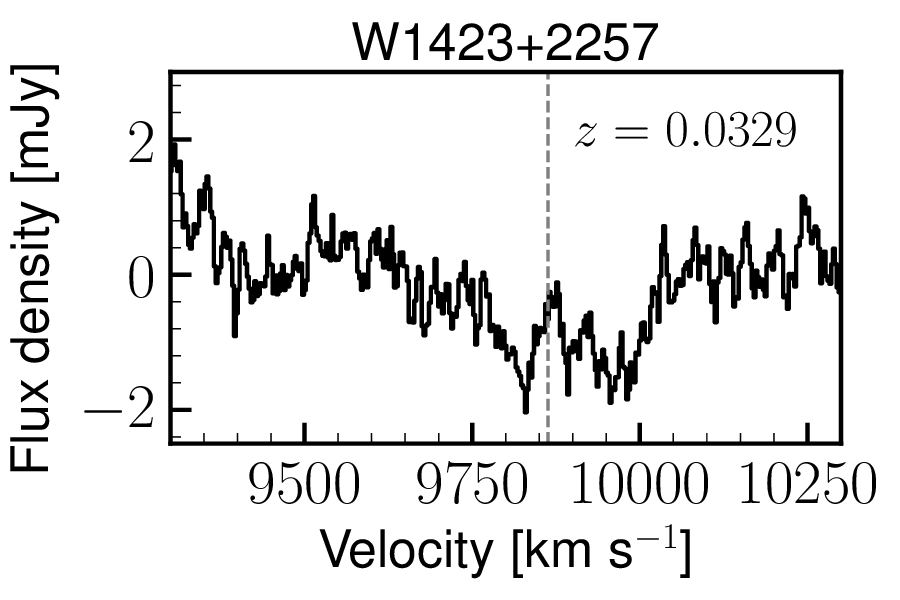}
	\includegraphics[scale=0.26]{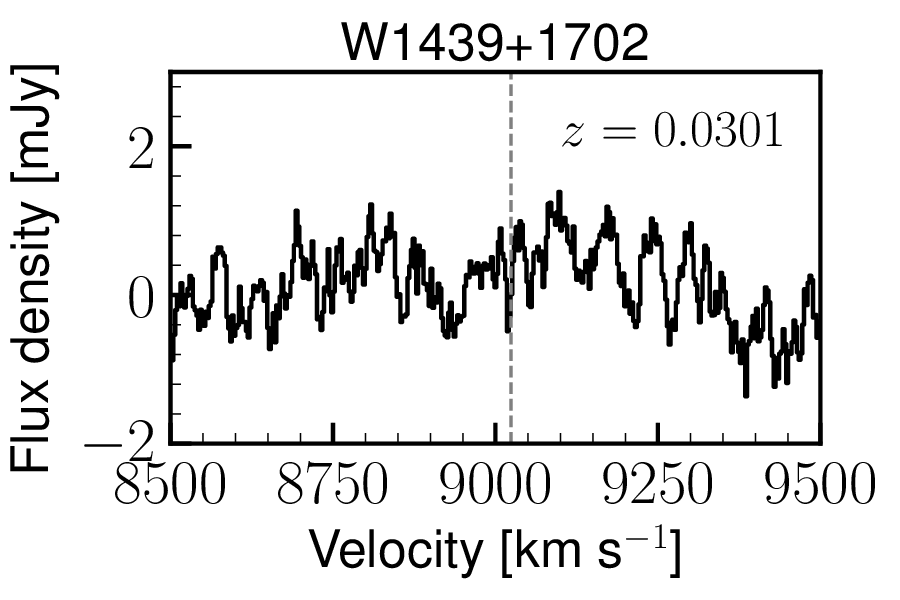}
	\includegraphics[scale=0.26]{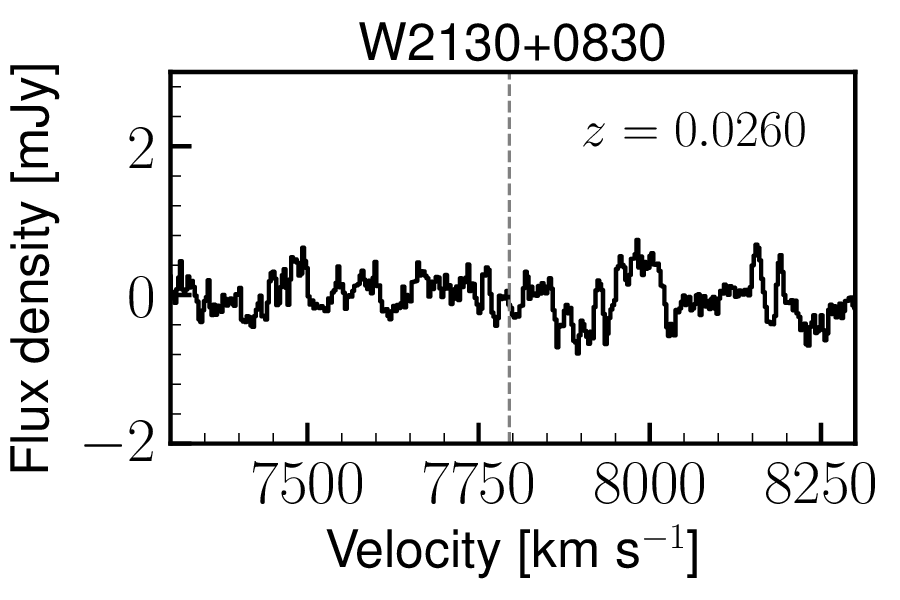}
	\includegraphics[scale=0.26]{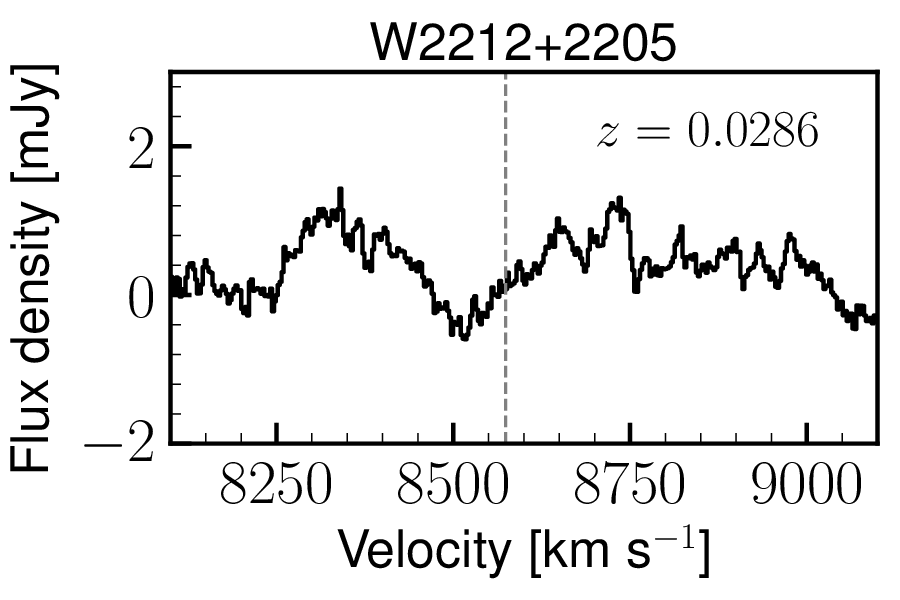}
	\includegraphics[scale=0.26]{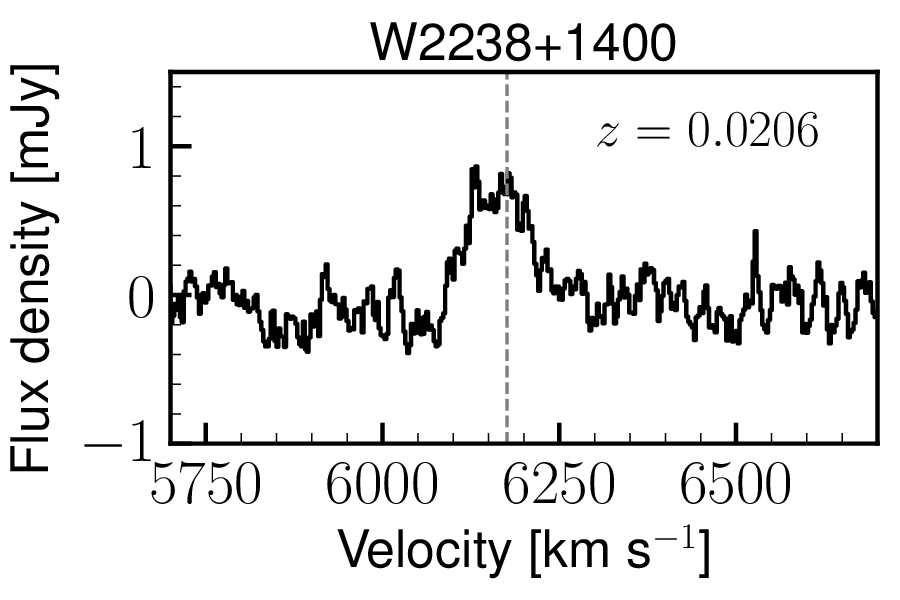}
	\includegraphics[scale=0.26]{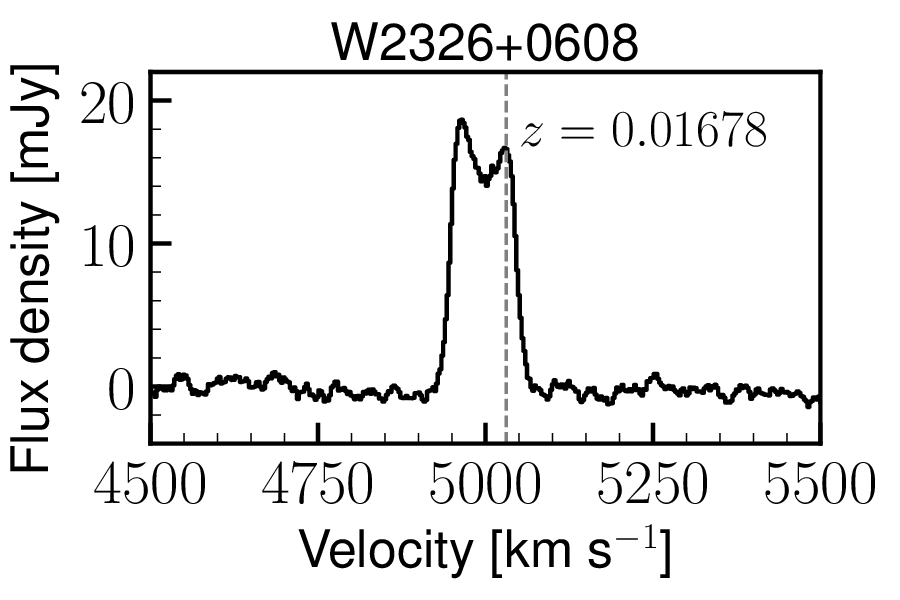}
	\caption{Arecibo H{\sc i} spectra of the 11 BCDs . Horizontal axes are velocities in the heliocentric frame of rest, and vertical axes are flux densities in mJy. H{\sc i} is detected in six BCDs: W0231+2441, W0801+2640, W0830+0225, W1016+3754, W2238+1400 and W2326+0608. The optical redshifts are shown at the upper right corner of the spectra. The systemic velocities from optical redshifts are marked with dashed vertical grey lines.}
	\label{profiles}
\end{figure*}

\subsection{Additional sources from literature}
\label{sec:litsample}
To compare  the properties of the 11 BCDs observed with the Arecibo, we collected information  on  several dIs/BCDs with  H{\sc i} observation  from the literature (see references in Tables~\ref{mirBCDs} and \ref{otherBCDs}). Sources with possible contamination by  gas-rich galaxy neighbours and H{\sc i} observations without sufficient spatial resolution have been excluded from  our analysis. If multiple H{\sc i} observations are available, measurements with better sensitivity and resolution are selected. We also did not include sources reported with H{\sc i} non-detections or uncertain detections. This resulted in a sample of additional 297 dIs/BCDs. After including the 11 BCDs observed with the Arecibo telescope, the total number of dIs/BCDs is 308. Although we excluded non-detections from the literature in our analysis due to non-uniform sensitivities, we included non-detections from our observations for completeness purposes, conservatively assuming the upper limit values as detections. This has no statistically  significant effect on our conclusions. Of these 308, we have both  stellar masses and star formation rates for 229 sources (also see Section \ref{sec:smsfroh}). We have based our analysis on these 229 sources. Of these 229, 225 sources have the data in AllWISE and  44 sources, including 11 sources observed with the Arecibo observatory, satisfy our MIR  selection criteria.
We also use the Arecibo Legacy Fast ALFA (ALFALFA) 100\% survey-SDSS cross-match  sample \citep{2020AJ....160..271D}, a sample of dwarf galaxies detected in  H{\sc i} by  ALFALFA survey \citep{2012AJ....143..133H}  and the extended GALEX Arecibo SDSS survey \citep[xGASS,][]{2018MNRAS.476..875C} for comparison. 
\section{Results}
\label{sec:results}
 H{\sc i} profiles from observations with the Arecibo are shown in Figure~\ref{profiles}. We detect H{\sc i} in six out of eleven sources in the sample. Of the six detections, four sources, W0801+0264, W0830+0225, W1016+3754, W2326+0608, have peak flux ($S_{\rm peak}$) above 5$\sigma$ significance.  W2238+1400, has  a 4$\sigma$ detection and W0231+2442 has  a marginal detection (2$\sigma$).  H{\sc i} mass has been calculated using the equation from \cite{1975gaun.book..309R}

\begin{equation}
 M_\mathrm{HI} \sim 2.36 \times 10^{5}\,\times \left(\frac{D_{\rm L}}{\rm Mpc}\right)^{2} \times \left(\frac{F_{\rm HI}}{\rm Jy\,km\,s^{-1}}\right){\rm M}_{\odot} , 
\label{eq1}
\end{equation} 
where $D_{\rm L}$ is the luminosity distance and 
\begin{eqnarray}
F_{\rm HI}=\int \frac{S(v)}{\rm Jy}\frac{{\rm d}v}{\rm km\,s^{-1}}
\end{eqnarray}
  is the integrated flux over the velocity range where the H{\sc i} line has been detected. As it is difficult to find a good distance estimator for these compact galaxies we have used luminosity distance for most of the sources. However, we have used the distances estimated using the tip of the red giant branch \citep[TRGB; ][]{lee1993ApJ...417..553L} or Tully-Fisher \citep[TF;][]{tf1977A&A....54..661T} method for the nearby sources (c*$z$ $\lesssim$ 500 km s$^{-1}$) where it is available in the literature. The median H{\sc i} mass is 10$^{8.3}$ M$_{\odot}$. 

The upper limit to the H{\sc i} mass was calculated assuming a 5$\sigma$ detection of a single peak boxcar spectral profile with velocity width at half peak flux of 50 km s$^{-1}$:
\begin{equation}
M_\mathrm{HI} \sim 2.36 \times 10^{5} \times \left(\frac{D_{\rm L}}{\rm Mpc}\right)^{2}\times 5\Delta S_\mathrm{rms}\Delta v \sqrt{\frac{50}{\Delta v}}    {\rm M}_{\odot},
\label{eq2}
\end{equation}
where $\Delta v$ is the velocity resolution in km s$^{-1}$ (10 km s$^{-1}$) and $\Delta S_\mathrm{rms}$ is the noise level per channel in Jy. The upper H{\sc i} mass limit for three sources (W1408+1753, W1423+2257, W2212+2205) was calculated using larger ``effective'' noise levels (0.6 mJy, 1.0 mJy, 0.8 mJy, respectively) than reported in Table~\ref{sourchar1} because of significant baseline structure present near the expected location of the source. 
The median upper limit on H{\sc i} mass for the non-detections is also estimated to be $10^{8.3}\,{\rm M}_{\odot}$.

 W1016+3754 was  previously observed in H{\sc i} by \cite{2007A&A...464..859P} using the Nancay Radio Telescope (NRT), though the H{\sc i} profile experienced confusion the with neighbouring galaxy UGC 5540 in the large Nancay beam. \cite{2007A&A...464..859P} derived H{\sc i} mass of $\log(M_\mathrm{HI}/{\rm M}_{\odot}$) $=$ 7.9 for  W1016+3754. We obtain  a slightly lower  H{\sc i} mass of $\log(M_\mathrm{HI}/{\rm M}_{\odot}$) $=$ 7.5 after resolving the object with Arecibo. W2238+1400 was  previously observed for H{\sc i} emission by \cite{2007A&A...464..859P} with NRT and \cite{2013A&A...558A..18F} with Effelsberg without a detection. We detect  H{\sc i}  after over two hours of integration with  Arecibo. Of the six detections, two sources (W0801+2640, W2326+0608) have double horn profile shapes while the other four (W0231+2441, W0830+0225, W1016+3754, W2238+1400) have single component profiles. W2326+0608 has a neighbouring galaxy that could not be resolved with the Arecibo beam. From high-resolution GMRT observations towards  W2326+0608, we find dense  ($\gtrsim$10$^{21}$ cm$^{-2}$)  H{\sc i} gas near the BCD \citep{chandola2023MNRAS.523.3848C}. The parameters  reported from  GMRT which we use in our analysis, are also provided in Table~\ref{sourchar1}.

Previous literature has noted that there is often a large amount of H{\sc i} gas outside the optical disk and the dwarf galaxies show a larger fraction of gas outside the optical disk \citep{wangjing2014MNRAS.441.2159W, wangjing2016MNRAS.460.2143W,wangjing2020ApJ...890...63W,hunt2020A&A...643A.180H}. For our sample, it is impossible to locate the H{\sc i} gas due to our limited spatial resolution. Hence, we use the global H{\sc i} mass for our study. This could create a bias if massive galaxies  have less gas outside the optical disk than less massive galaxies. To check this, we follow the approach taken by \cite{hunt2020A&A...643A.180H}. We find that the optical-to-H{\sc i} radius ratio and the ratio of H{\sc i} mass outside the optical disk to total H{\sc i} mass do not change systematically with stellar mass for our sample, implying that the sample is unbiased (see Section~\ref{hilocation}).  

\section{Stellar mass ($M_{\ast}$)  and star formation rates (SFR)}
\label{sec:smsfroh}
\begin{figure*}
	\center
	\includegraphics[scale=0.6]{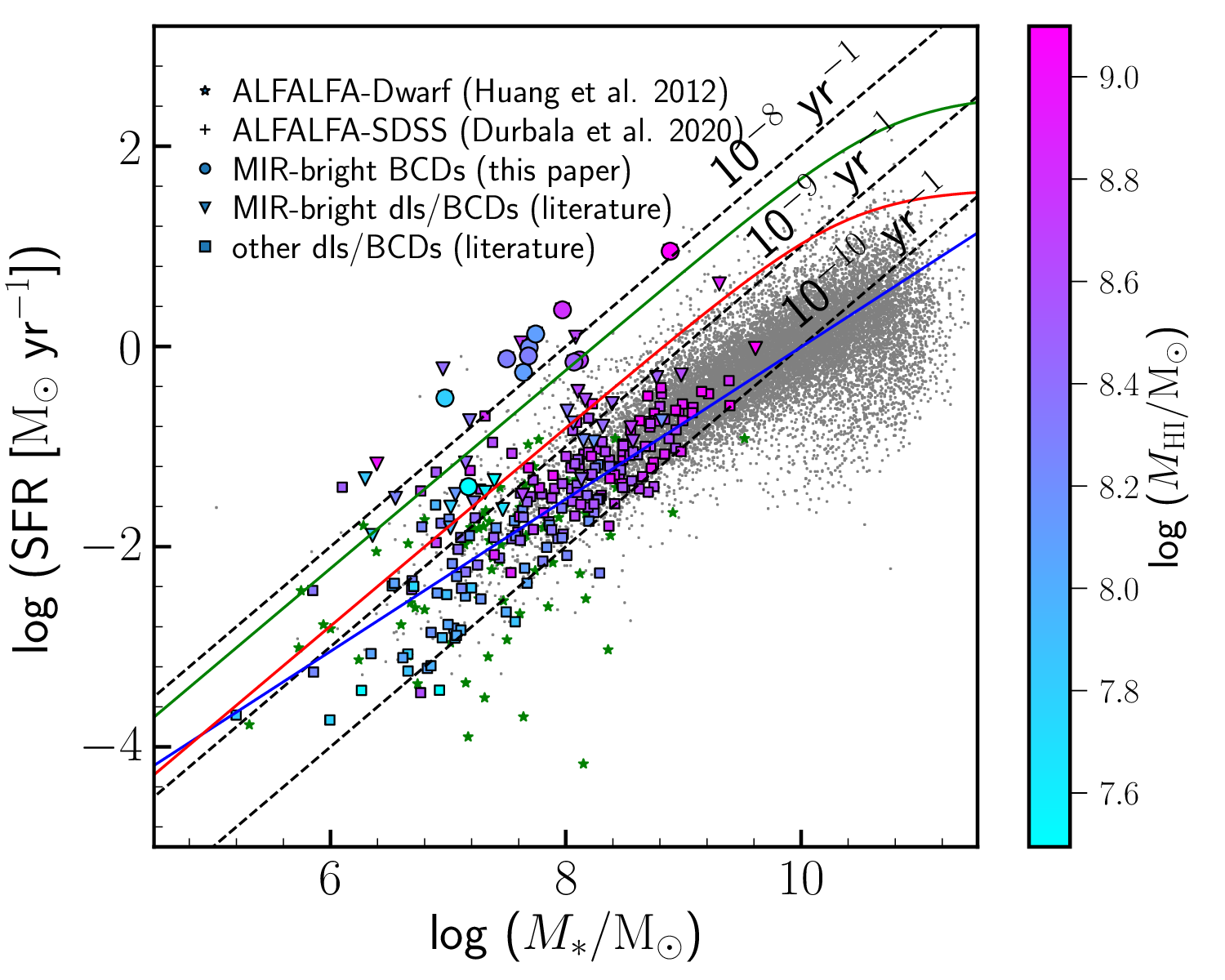}
	\caption{SFR vs  M$_{\ast}$. Different symbols mean the same as in Fig.~\ref{selectioncriteria}. The colour scale shows the H{\sc i} mass of the dIs/BCDs. The solid lines in blue, red and green colours show fits to the main sequence star-forming galaxies at $z$= 0 \citep{2015ApJ...801L..29R}, 1 and  6 \citep{popesso2023MNRAS.519.1526P},  respectively. Dashed diagonal lines in black show  different sSFR levels. Most of the BCDs observed with the Arecibo  have their sSFRs higher than the main sequence star-forming galaxies at $z$=6.  Data from ALFALFA(100 \%)-SDSS survey \citep{2020AJ....160..271D} is also shown in grey coloured \lq +' symbols. Stellar masses for the ALFALFA-SDSS sample were derived using the method of \protect\cite{2011MNRAS.418.1587T}. SFRs were derived from \textit{GALEX} NUV luminosities corrected for extinction  using  MIR 22 \micron  \citep {2020AJ....160..271D}. ALFALFA dwarf galaxies \citep{2012AJ....143..133H} are also shown in green.}
	\label{SFRstellarmass}
\end{figure*}
To  estimate the properties of the
dIs/BCDs, we assembled the multi-band photometric data from \textit{Galaxy Evolution Explorer} \citep[\textit{GALEX;}][]{2017ApJS..230...24B}, Panoramic Survey Telescope and Rapid Response System  \citep[Pan-STARRS;][]{flewelling2020ApJS..251....7F}, SDSS \citep{2015ApJS..219...12A}, Two Micron All Sky Survey\citep[2MASS;][]{2006AJ....131.1163S}, \textit{UKIRT} Infrared Deep Sky Survey\citep[UKIDSS;][]{2007MNRAS.379.1599L} and AllWISE \citep{2014yCat.2328....0C} to construct the UV to Infrared spectral energy distribution (SED). We used a cross-match radius of 5 arcseconds around SDSS/Pan-STARRS optical coordinates to obtain photometric data points for the SED. To decompose the stellar components in the host galaxy, we use the method of \cite{2008ApJ...676..286A,2010ApJ...713..970A} in which the SED of a galaxy is  described by the   linear contribution of empirical spectral templates \citep{2010ApJ...713..970A}.  The three empirical SED templates -- an elliptical galaxy (E),  a spiral galaxy (Sbc), and  an irregular galaxy (Im)-- are used to model the SED of the host galaxy in the wavelength range  0.03-30 \micron  \citep{2010ApJ...713..970A}. The E, Sbc and Im templates represent components of an old stellar population, a continuously star-forming population and a starburst population, respectively.

\subsection{Stellar masses}
 Based on the best model  for each SED, we derived the K-correction. Then using the correlations between rest frame $ugriz$ colours and M/L ratios reported by \cite{2003ApJS..149..289B}, we estimated the stellar mass-to-light ratio for the host galaxies. Using this method, we obtained  stellar mass estimates for 214 sources. We compared the stellar mass of sources common in  \textit{GALEX-}SDSS-\textit{WISE} Legacy Catalog  \citep[GSWLC,][]{salim2016ApJS..227....2S,salim2018ApJ...859...11S} and our compilation of dIs/BCDs using M/L from \protect\cite{2003ApJS..149..289B}. We have 28 sources common in both samples with no troublesome photometry in GSWLC. As \cite{2003ApJS..149..289B} consider a relatively smooth star-forming history and have higher estimates of M/L\citep{zibetti2009MNRAS.400.1181Z}, we find stellar mass estimates are slightly larger by $\sim$0.1 dex compared to the method used by \cite{salim2016ApJS..227....2S,salim2018ApJ...859...11S}.   For another 11 sources, we obtained stellar masses from the SDSS MPA-JHU database \citep{2004astro.ph..6220B} \footnote{\url{http://wwwmpa.mpa-garching.mpg.de/SDSS/DR7/Data/stellarmass.html}} where stellar masses were  derived using fits to the SDSS photometry. For another four sources, we use  values from the literature (see the references in Tables ~\ref{mirBCDs} and ~\ref{otherBCDs}). 

\subsection{SFRs and sSFRs}
Considering  extinction from  dust in the host galaxies, we correct the \textit{GALEX}  FUV and  NUV luminosity using the \textit{WISE} W4 , and estimate the extinction-corrected SFR using the method of  \cite{2011ApJ...741..124H} and \cite{2012ARA&A..50..531K}. We estimate the extinction-corrected FUV/NUV SFR for 148 galaxies. For another 21 galaxies, where the FUV/NUV fluxes are not available from \textit{GALEX}, we predict the FUV luminosities using the model SED fitting and correct for extinction using the W4 values.  

 Some BCDs are young starburst galaxies with little or no dust extinction \citep{2007ApJ...662..952W}. For 53 such sources, where there is no clear detection in  the \textit{WISE}  W4 band, we use FUV/NUV SFR without correcting for dust extinction. For another 5 sources, we use the SFR from MPA-JHU catalog \citep{2004astro.ph..6220B} \footnote{\url{http://wwwmpa.mpa-garching.mpg.de/SDSS/DR7/sfrs.html}} where the SFR with corrections for dust attenuation were  derived  using the methods of \cite{2004MNRAS.351.1151B} and aperture corrections were applied using the methodology of \cite{2007ApJS..173..267S}. For 2 additional galaxies, we obtained  values from the literature.

 In Fig.~\ref{SFRstellarmass}, we plot star formation rate (SFR)  versus stellar mass ($M_{\ast}$) for the sample of 229 dIs/BCDs (44 MIR-bright dIs/BCDs). SFRs and sSFRs  for MIR-bright dIs/BCDs are $\sim$10 times higher than typical values for other dIs/BCDs of similar stellar masses. The median SFR is $10^{-0.74}\,{\rm M}_{\odot}\,{\rm yr}^{-1}$  for MIR-bright BCDs  as compared to $10^{-1.46}\,{\rm M}_{\odot}$ yr$^{-1}$ for other dIs/BCDs. The median sSFR values are $\sim$ 10$^{-8.6}$ yr$^{-1}$ and $\sim$10$^{-9.5}$ yr$^{-1}$ for MIR-bright and other dIs/BCDs, respectively. We also show  fits to  main sequence star-forming galaxies ($M_{\ast} >$10$^{10}$ M$_{\odot}$) at $z=$0 \citep{2015ApJ...801L..29R}, $z=1$ and $z=6$ \citep{popesso2023MNRAS.519.1526P}  in  Fig.~\ref{SFRstellarmass}. The median sSFR (10$^{-7.8}$ yr$^{-1}$) for MIR-bright BCDs observed with Arecibo is higher (sSFR$>$ 10$^{-8}$ yr$^{-1}$) than main sequence star-forming galaxies at $z=$6. In the fundamental formation relation \citep[FFR;][]{2021ApJ...907..114D, 2021ApJ...915...94D}, sSFR can be expressed as
 	\begin{equation}
 		{\rm sSFR}=f_{\rm HI} \times {\rm SFE}
 	\end{equation}
 	where
 	$f_{\rm HI}=$$M_{\rm HI}$/$M_{\ast}$ is the gas fraction or H{\sc i} mass to stellar mass ratio and SFE is the star formation efficiency of the gas. SFE  is given by the equation
 	\begin{equation}
 		{\rm SFE}=\frac{\rm SFR}{M_{\rm HI}}=\frac{1}{\tau_{\rm HI}}
 	\end{equation}
 	where $\tau_{\rm HI}$ is the atomic gas depletion timescale.
 The  high sSFR for these objects could be  due to higher gas fractions or higher star formation efficiencies (lower depletion timescales) of the atomic H{\sc i} gas,  or both.    Table~\ref{medianval} reports the median values of $M_{\ast}$, sSFR, $f_{\rm HI}$ and $\tau_{\rm HI}$ for different subsamples of $M_{\ast}$ and sSFR.   Table~\ref{correl} reports the Spearman's correlation coefficients of $f_{\rm HI}$ and $\tau_{\rm HI}$ relation with $M_{\ast}$ and sSFR. In the next section, we discuss the H{\sc i} mass to stellar mass ratio and gas depletion timescale of these galaxies  to understand their  high sSFRs.
 
\section{Analysis and discussion}
\label{sec:analysis}
\begin{figure*}
	\center
	
	\includegraphics[scale=0.55]{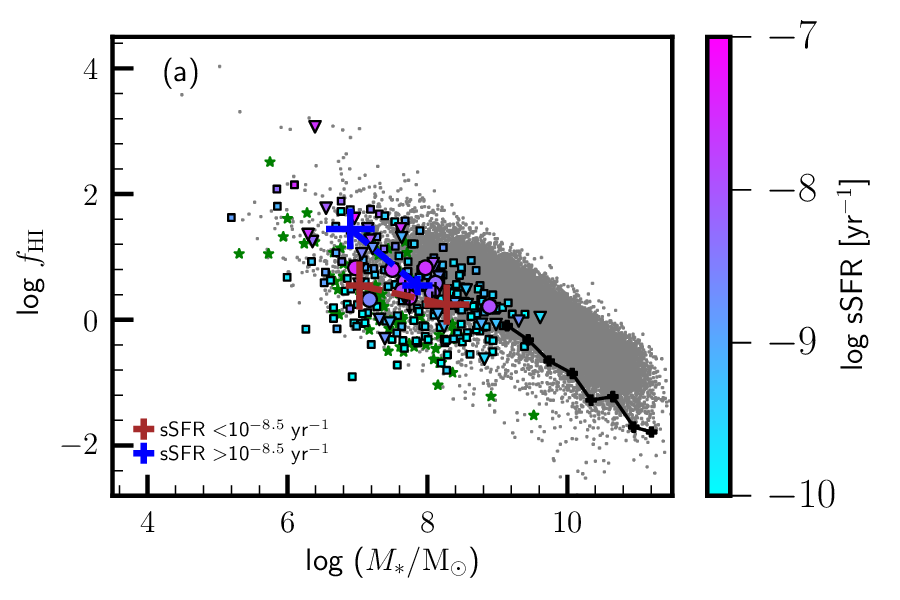}
	\includegraphics[scale=0.55]{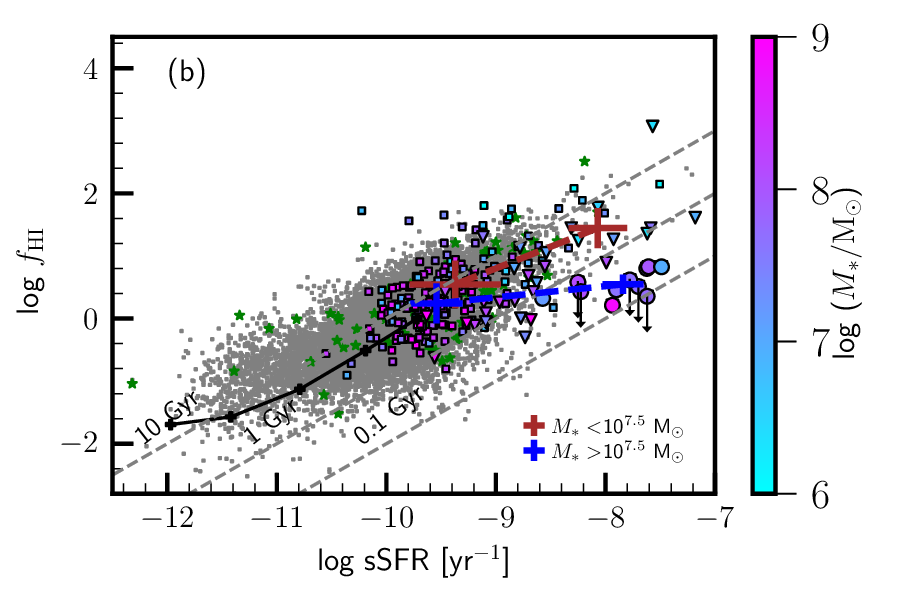}
	\includegraphics[scale=0.55]{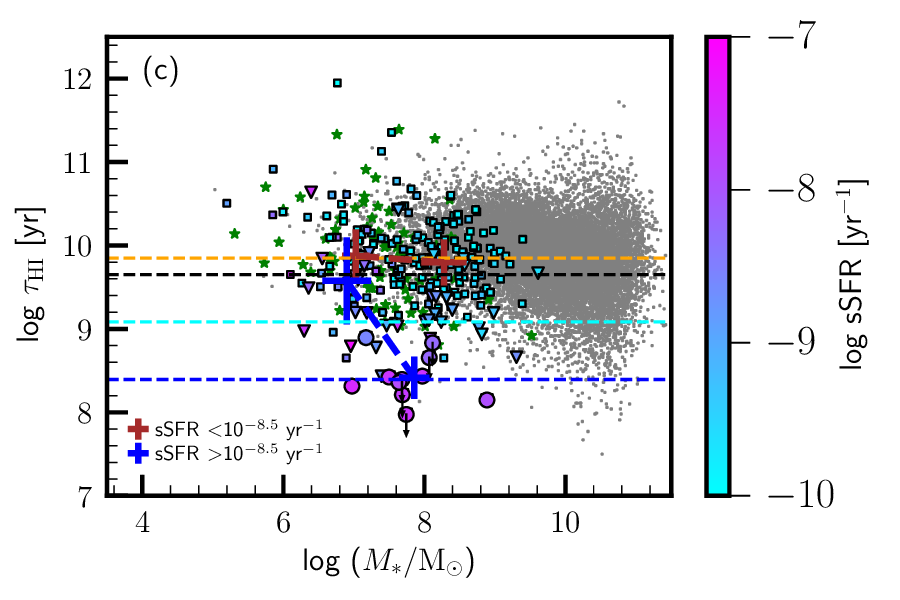}
	\includegraphics[scale=0.55]{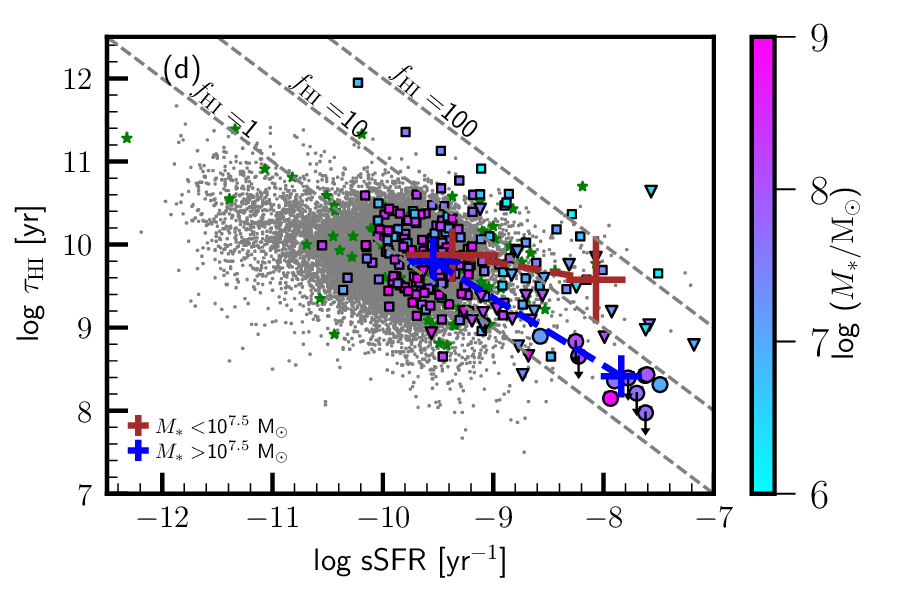}		
	\caption{(a) $f_{\rm HI}$ vs. $M_{\ast}$ for all subsamples of dIs/BCDs.  Symbols  are the same as in Fig.~\ref{SFRstellarmass}. Median values of $f_{\rm HI}$ for different subsamples of stellar masses and sSFR are shown with blue (sSFR $>$ 10 $^{-8.5}$ yr$^{-1}$) and brown (sSFR $\leq$ 10$^{-8.5}$ yr$^{-1}$) colour error bars. (b) $f_{\rm H \sc{I}}$ vs. sSFR.  Median values of $f_{\rm HI}$ for different subsamples of stellar masses and sSFR are shown with blue ($M_{\ast}>10^{7.5}\,{\rm M}_{\odot}$) and brown ($M_{\ast}\leq 10^{7.5}\,{\rm M}_{\odot}$) colour error bars. Dashed diagonal grey lines show the depletion times scales. In plots (a) and (b) median $f_{\rm HI}$ for  xGASS galaxies \citep{2018MNRAS.476..875C} are shown in black colour. (c) Gas depletion timescale ($\tau_{\rm HI}$) vs. $M_{\ast}$.  Dashed horizontal lines represent the median of $\tau_{\rm HI}$ for other dIs/BCDs in orange,  all MIR-bright dIs/BCDs in cyan and MIR-bright BCDs observed with the Arecibo observatory in blue. Black dashed horizontal line represents the median $\tau_{\rm HI}$ for local main sequence galaxies ($M_{\ast}> 10^{9}\,{\rm M}_{\odot}$) from \protect\cite{2017ApJS..233...22S} and \protect\cite{2018MNRAS.476..875C}.  Median values of $\tau_{\rm HI}$ for different subsamples of stellar masses and sSFR are shown with blue (sSFR $>$ 10$^{-8.5}$ yr$^{-1}$) and brown (sSFR $\leq$10$^{-8.5}$ yr$^{-1}$) colour error bars. (d) $\tau_{\rm HI}$ vs.  sSFR.  Median values of $\tau_{\rm HI}$ for different subsamples of stellar masses and sSFR are shown in blue ($M_{\ast}>10^{7.5}\,{\rm M}_{\odot}$) and brown ($M_{\ast} \leq 10^{7.5}\,{\rm M}_{\odot}$) colour error bars.  The dashed grey diagonal lines represent  different levels of $f_{\rm HI}$. For $f_{\rm HI}=$1, the depletion time scale is equal to the e-folding time scale. In plots (a) and (c) colour scales show sSFR, while in plots (b) and (d) it shows the stellar masses.}
	\label{ffr}
\end{figure*}

\begin{figure}
	\center
	\includegraphics[scale=0.55]{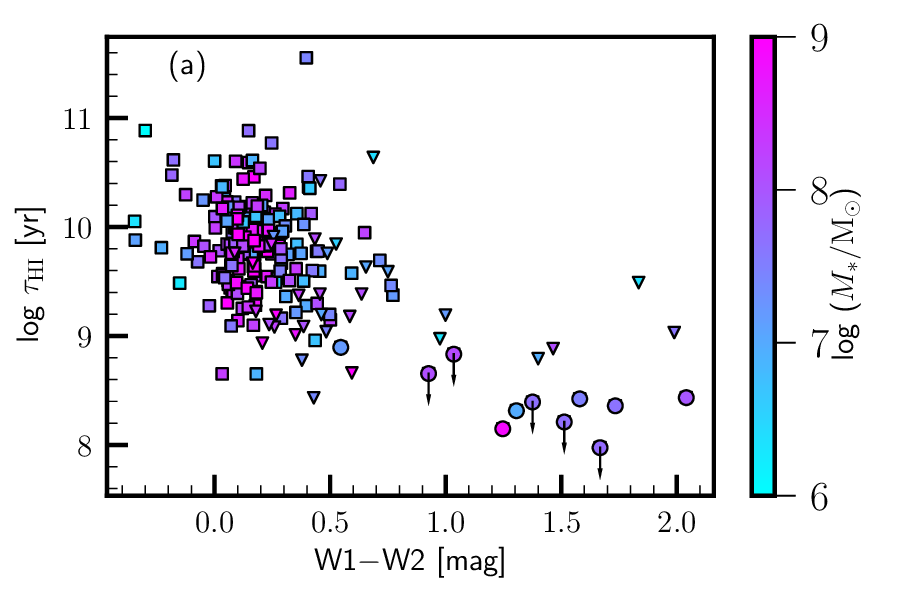}
	\includegraphics[scale=0.55]{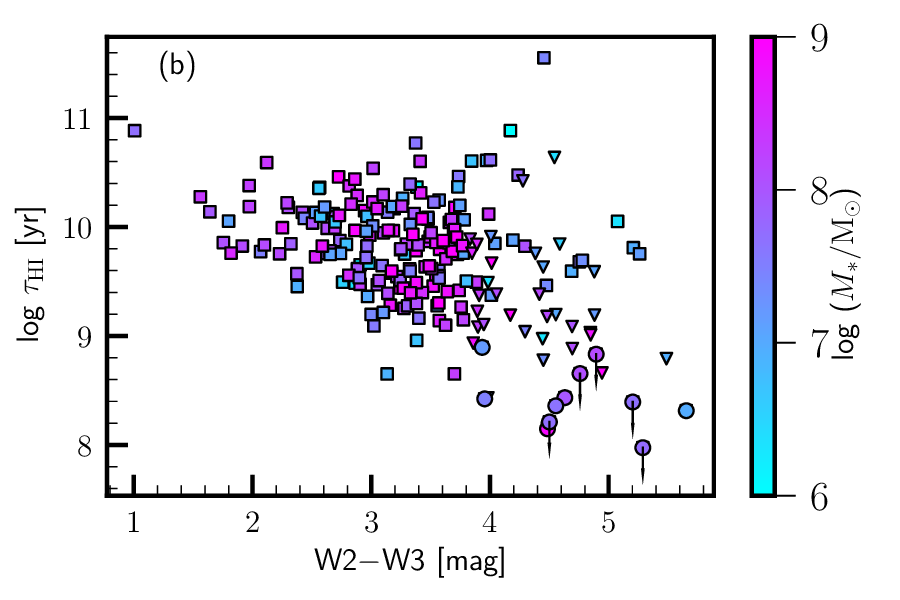}

	\caption{$\tau_{\rm HI}$ vs \textit{WISE} W1$-$W2 and  \textit{WISE} W2$-$W3 in panels (a) and (b) respectively. Symbols mean the same as in Fig.~\ref{selectioncriteria}. The Arecibo-observed BCDs with young starburst (sSFR $\gtrsim$ 10$^{-8.5}$ yr$^{-1}$) and low depletion time scales show very red W1$-$W2 and W2$-$W3 colours.}
	\label{wisecolor}
\end{figure}

\begin{table*}
	\begin{center}
		\caption{Median values of $M_{\ast}$, sSFR, $f_{\rm HI}$ and $\tau_{\rm HI}$ for different subsamples of $M_{\ast}$ and sSFR.}
		\label{medianval}
		\begin{tabular}{ccccccc}
			\hline
			$M_{\ast}$ category & sSFR category & Median log $M_{\ast}$  & Median log sSFR  & Median log $f_{\rm HI}$  & Median log $\tau_{\rm HI}$& N$^{\dagger}$ \\
			&  & [${\rm M}_{\odot}$]  & [yr$^{-1}$]  &  & [yr]&  \\
			\hline
		
			low ($\leq$10$^{7.5}$M$_{\odot}$)& low ($\leq$10$^{-8.5}$yr$^{-1}$) & 7.0$\pm$0.2 & $-$9.4 $\pm$0.4 & 0.5$\pm$0.4 & 9.9$\pm$0.3 &58\\
		
			low ($\leq$10$^{7.5}$M$_{\odot}$) & high ($>$10$^{-8.5}$yr$^{-1}$) & 6.9$\pm$0.3 & $-$8.1$\pm$0.3 & 1.4$\pm$0.3 & 9.6$\pm$0.5& 17 \\
		
			high ($>$10$^{7.5}$M$_{\odot}$) &low ($\leq$10$^{-8.5}$yr$^{-1}$) & 8.3$\pm$0.3 & $-$9.5$\pm$0.2 & 0.2$\pm$0.3 & 9.8$\pm$0.3 & 144 \\
		
			high ($>$10$^{7.5}$M$_{\odot}$)  & high ($>$10$^{-8.5}$yr$^{-1}$) & 7.9$\pm$ 0.2 & $-$7.8$\pm$0.2 & 0.5$\pm$0.2 & 8.4$\pm$0.3 & 10\\
			\hline
		\end{tabular}
	\end{center}
	$^{\dagger}$Number of sources.
\end{table*}
\begin{table}
	\begin{center}
		\caption{Spearman's correlation coefficients (R-value) of $f_{\rm HI}$ and $\tau_{\rm HI}$ relation with $M_{\ast}$ and sSFR for different subsamples of $M_{\ast}$ and sSFR.}
		\label{correl}
		\scriptsize{
			\begin{tabular}{cccccc}
				\hline
				Subsample & X-axis & Y-axis  & R-value  & p-value  & N$^{\dagger}$ \\
				\hline
				sSFR  $\leq$10$^{-8.5}$yr$^{-1}$ &  $M_{\ast}$ & $f_{\rm HI}$  &$-$0.4 &1.1$\times$10$^{-09}$ & 202\\
				
                sSFR $>$10$^{-8.5}$yr$^{-1}$ & $M_{\ast}$  & $f_{\rm HI}$  &$-$0.73 & 0.0021 & 27 \\
				
				sSFR $\leq$10$^{-8.5}$yr$^{-1}$  & $M_{\ast}$  & $\tau_{\rm HI}$  & $-$0.13 & 0.06 & 202\\
				
    sSFR $>$10$^{-8.5}$yr$^{-1}$ & $M_{\ast}$  & $\tau_{\rm HI}$  &$-$0.63  & 0.012 & 27\\
				
				$M_{\ast}$$\leq$10$^{7.5}$M$_{\odot}$ &  sSFR & $f_{\rm HI}$  & 0.63  &3.3$\times$10$^{-5}$ & 75\\
				
				$M_{\ast}$$>$10$^{7.5}$M$_{\odot}$&  sSFR & $f_{\rm HI}$& 0.56 & 4.34$\times$10$^{-17}$ & 154\\
				
				$M_{\ast}$$\leq$10$^{7.5}$M$_{\odot}$&  sSFR & $\tau_{\rm HI}$  &$-0.34$ &0.04 & 75\\
			
				$M_{\ast}$$>$10$^{7.5}$M$_{\odot}$&  sSFR & $\tau_{\rm HI}$  &$-$0.41  &2.5$\times$10$^{-9}$ &154 \\
				
				\hline
		\end{tabular}}
		$^{\dagger}$Number of sources.
	\end{center}
	
\end{table}

\subsection{H{\sc i} mass to stellar mass ratio ($f_{\rm HI}= M_{\rm HI}/M_{\ast}$ )}

 Fig.~\ref{ffr}  shows $f_{\rm HI}$ vs. $M_{\ast}$ and $f_{\rm HI}$ vs. sSFR for all subsamples mentioned earlier. The $f_{\rm HI}$ values vary between $\sim$0.01 and 1000 for different stellar masses of these sources. Median $f_{\rm HI}$ values are 10$^{0.5\pm0.4}$($\sim$3.2) and 10$^{0.3\pm0.4}$($\sim$2.0) for MIR-bright and other dIs/BCDs, respectively. The median $f_{\rm HI}$ for the Arecibo observed sample is 10$^{0.5\pm0.2}$ ($\sim$3.2).  The $f_{\rm HI}$  vs. $M_{\ast}$ plot in  panel (a) shows a negative correlation between the two parameters for all subsamples. $f_{\rm HI}$ increases gradually with  decreasing  $M_{\ast}$, which is consistent with the findings of \cite{2018MNRAS.476..875C}. This suggests that the baryonic mass is dominated by cold atomic gas at lower stellar masses. This anti-correlation is stronger for higher sSFR ($>$10$^{-8.5}$ yr$^{-1}$) dIs/BCDs as compared to lower sSFR  dIs/BCDs ($<$10$^{-8.5}$ yr$^{-1}$). At lower stellar masses ($<$10$^{7.5}$ M$_{\odot}$) dIs/BCDs with higher sSFR have higher gas fractions. MIR-bright as well as the other dIs/BCDs follow the same relation implying there is no difference in the baryonic composition of MIR-bright and other dIs/BCDs at  a given stellar mass. 

 Large area shallow H{\sc i} surveys such as  ALFALFA are biased towards the gas-rich galaxies of different stellar masses due to their sensitivity limitation, and hence we see a higher density of ALFALFA-SDSS sources towards the higher values of $f_{\rm HI}$ in  $f_{\rm HI}$ vs. $M_{\ast}$ relation. This is not the case for targeted  samples selected based on  optical criteria where gas-poor sources are also detected with  $M_{\rm HI}$/$M_{\ast}$ ratio up to $\sim$2 \% \citep{2018MNRAS.476..875C}. The median $f_{\rm HI}$ for local star-forming main sequence galaxies with an average stellar mass of 10$^{9.14}$ from xGASS sample 
 \citep{2018MNRAS.476..875C}  is  $\sim$0.8. This is a factor of 3 lower than the median values for dIs/BCDs in the local universe, implying a strong dependence of $f_{\rm HI}$ on stellar mass. From stacking  starforming galaxies at an average redshift of $\sim$1.0, a recent study by \cite{chowdhury2022ApJ...937..103C} finds that $f_{\rm HI}$ is $\sim$1.4. Hence, dIs/BCDs in the local universe have higher $f_{\rm HI}$ than the $z \sim$1 starforming main sequence massive galaxies ($>$10$^{9}$ M$_{\odot}$) as well. 
 
 Fig.~\ref{ffr} (b) shows $f_{\rm HI}$ as function of sSFR. For both types of dIs/BCDs, the $f_{\rm HI}$ increases with increasing sSFR. However, at higher sSFR ($>$10$^{-8.5}$ yr$^{-1}$) there is scatter in $f_{\rm HI}$  due to different stellar masses. The dIs/BCDs with lower stellar masses have a higher median $f_{\rm HI}$ (10$^{1.4\pm0.3}$). The dIs/BCDs with higher sSFR and higher stellar masses appears to diverge from the $f_{\rm HI}$-sSFR relation  towards lower median $f_{\rm HI}$ (10$^{0.5\pm0.2}$). The MIR-bright BCDs observed with the Arecibo  have the lower  $f_{\rm HI}$ as compared to the dIs/BCDs of similar sSFR studied in the literature. This implies less fuel per unit stellar mass is available for a similar star-formation rate per unit stellar mass  and hence, a lesser role of $f_{\rm HI}$ in the current starburst episode.

\subsection{Atomic H{\sc i} gas depletion timescales}
 Cold neutral gas is considered  the fuel reservoir for star formation activity, and the global H{\sc i} content of galaxies is  related to how long the current star formation activity can be supported. 
We estimate the H{\sc i} gas depletion time scale  ($\tau_{\rm HI}$) using 
\begin{equation}
\tau_{\rm HI}  = \frac{1}{\rm SFE}= \frac{M_{\rm HI}}{\rm SFR} = \frac{f_{\rm HI}}{\rm sSFR}. 
\end{equation}
The median $\tau_{\rm HI}$ for MIR-bright and other dIs/BCDs are 10$^{9.1 \pm 0.4}$ yr  ($\sim$1.3 Gyr) and 10$^{9.8\pm 0.3}$ yr ($\sim$6.3 Gyr), respectively.  At the current SFR, MIR-bright dIs/BCDs will  run out of their fuel earlier than other dIs/BCDs.  We find the median H{\sc i} atomic gas depletion timescales for the sample of  MIR-bright and other dIs/BCDs are $\sim$0.26 and $\sim$1.26 times $\sim$5 Gyr, the gas depletion time scales of  higher stellar mass main sequence galaxies\citep{2017ApJS..233...22S}. If we only consider the BCDs observed with the Arecibo, the median $\tau_{\rm HI}$ is estimated to be 10$^{8.4 \pm 0.2}$ yrs or $\sim$0.3 Gyr (conservatively assuming the upper limits as detections). This is lower than $\sim$0.6 Gyr reported for relatively higher redshift Green Pea galaxies \citep{2021ApJ...913L..15K}. This is  more than an order of magnitude ($\sim 3.9\,\sigma$) lower than  other dIs/BCDs. This is also lower than $\sim$1.7 Gyr,  the depletion timescale for high redshift  main sequence galaxies \citep[$z \sim 1$;][]{chowdhury2022ApJ...937..103C}. 

We examine the dependence of $\tau_{\rm HI}$ on $M_{\ast}$ and sSFR in the bottom panels of Fig.~\ref{ffr}. It appears that there is no significant correlation  (Spearman's R-value of $-$0.08, p-value of 0.21)  between $\tau_{\rm HI}$ and $M_{\ast}$ for the combined sample of  dIs/BCDs (Fig.~\ref{ffr}c). There is a dispersion in the relation  because lower sSFR ($<$10$^{-8.5}$ yr$^{-1}$)  galaxies have higher  depletion timescales at higher stellar masses. Dividing the sample into lower and higher sSFR strengthens the anti-correlation  (Spearman's R-value: $-$0.63, p-value: 0.01) for higher sSFR objects.  Fig.~\ref{ffr}(d)  also shows a weak but significant negative correlation of $\tau_{\rm HI}$ with sSFR (Spearman's R value:$-$0.38, p-value: 2.7$\times$10$^{-9}$ ). Again in this relation, there is a dispersion due to  lower mass galaxies  having higher $\tau_{\rm HI}$ at higher sSFR ($>$10$^{-8.5}$yr$^{-1}$). There is stronger $\tau_{\rm HI}$-sSFR anti-correlation (R-value: $-$0.41, p-value: 2.5$\times$10$^{-9}$) for galaxies  with higher stellar mass ($>$10$^{7.5}$ M$_{\odot}$). MIR-bright dIs/BCDs from the Arecibo observations cover the highest sSFR and lowest $\tau_{\rm HI}$ region.

In the literature, almost no dependence of $\tau_{\rm HI}$ on $M_{\ast}$ was found by \cite{2017ApJS..233...22S}  for relatively higher mass galaxies ($>$10$^{9}$ M$_{\odot}$) and \cite{2016MNRAS.463.4268T} for low stellar mass ($\gtrsim$ 10$^{5}$ M$_{\odot}$) and low metallicity ($<$ 1/10 Z$_{\odot}$) BCDs.
However,  H{\sc i} selected samples from   the ALFALFA survey \citep[]{2011AJ....142..170H}  find that the depletion time scale decreases or star-formation efficiency, SFE=1/$\tau_{\rm HI}$, increases with stellar mass \citep{2012AJ....143..133H, 2012ApJ...756..113H, 2015ApJ...808...66J}, albeit subject to considerable scatter in $\tau_{\rm HI}$ at lower stellar masses.  Surveys like ALFALFA are likely to detect more H{\sc i} rich but optically faint objects compared to optically selected samples, and hence objects with higher depletion time scale (lower SFE) are only detected at low stellar masses \citep{2012AJ....143..133H,2012ApJ...756..113H}. Recently, \cite{hunt2020A&A...643A.180H} found that $\tau_{\rm HI}$ shows inverse relation with sSFR. Compared to \protect\cite{2012ApJ...756..113H}, \cite{2015ApJ...808...66J} and \cite{hunt2020A&A...643A.180H}, we explored higher sSFR regimes and find a much  stronger anti-correlation of $\tau_{\rm HI}$ with $M_{\ast}$ for higher sSFR, and with sSFR for higher stellar mass dwarf galaxies. 

 An almost constant atomic gas depletion timescale with $M_{\ast}$  and the sSFR may imply a balance between the gas accretion from intergalactic medium (IGM) and its conversion to molecular gas and finally to stars is regulated by external mechanisms \citep{2010MNRAS.408..919S}. In some earlier studies with optically selected samples, while atomic gas depletion timescales were found to be almost constant with $M_{\ast}$ and sSFR, molecular gas depletion timescales were found to vary with $M_{\ast}$ and sSFR, differing by more than an order of magnitude with atomic gas depletion timescales for lower stellar masses and high sSFRs \citep{2015A&A...583A.114H, 2017ApJS..233...22S}. These results were interpreted as a large fraction of H{\sc i} gas being not directly related to the star-formation process in these systems \citep{2017ApJS..233...22S}. Contrary to this, an inverse relation may imply a relatively larger fraction of H{\sc i} gas being actively involved in the current high star-burst episode. Since most of the sources have  depletion timescales larger than 1/sSFR, it should be possible to sustain the current SFR to  at least double  the stellar mass.

\subsection{Why does the Arecibo sample has the highest specific star formation rate or lowest depletion timescale?}
As mentioned earlier in the previous section, the high sSFR or starburst nature of the galaxies depends mainly on two factors: (1) the gas fraction ($f_{\rm HI}$ or availability of H{\sc i} gas per unit stellar mass), and (2) SFE  or SFR  per unit mass of H{\sc i} gas. For the starburst dIs/BCDs with sSFR $>$10$^{-8.5}$yr$^{-1}$,  the role of $f_{\rm HI}$ or $\tau_{\rm HI}$ in the current high starburst episode is strongly dependent on their stellar masses. At the lower stellar masses, it is mainly  $f_{\rm HI}$ which drives the starburst, while for higher stellar masses it is mainly SFE   of the gas. Based on  their stellar masses, $f_{\rm HI}$ of the MIR-bright BCDs in the Arecibo sample are  similar to the other low sSFR dIs/BCDs  in the literature. Hence, it is mainly the low gas depletion time scales or high SFE of the gas which drives the high sSFR or current starburst episode. In this subsection, we discuss the reason for the high depletion time scale in these systems. 
Many compact starburst BCDs with warm dust can be detected using \textit{WISE} colours \citep{2016ApJ...832..119H}. Hence, we plot the sample of dIs/BCDs in $\tau_{\rm HI}$ vs \textit{WISE } colour plots (Fig~\ref{wisecolor}a and Fig~\ref{wisecolor}b). Galaxies with lower depletion timescale have the redder W1$-$W2 and W2$-$W3 colours. The sources in the Arecibo sample have the reddest W1$-$W2 colours, indicating the presence of warm dust. In the sample  of  11 BCDs observed with Arecibo, we find that 10 BCDs have W1$-$W2 $>$ 0.8. Note that this red colour is not due to active galactic nuclei (AGN), which have higher luminosities \citep{2016ApJ...832..119H}.  Among  all dIs/BCDs, 16 sources satisfy this criterion. The median H{\sc i} depletion time scale for the dIs/BCDs with red W1$-$W2 colours is $\sim$0.46 Gyr. BCDs with redder W1$-$W2 colours are known to have larger H$\alpha$ equivalents widths (EQW), H$\alpha$ luminosities and ionizing photon rates \citep{2016ApJ...832..119H}. The similar gas fraction ($f_{\rm HI}$) at a  given stellar mass suggests  that despite the high ionizing photon rate there is no change in the baryonic composition of  the Arecibo sample. This could be due to the high star-forming gas density shielding  H{\sc i} gas from  ionization \citep{2004A&A...421..555H}. This high density of gas could also create the favourable conditions for faster conversion of H{\sc i} to H$_{2}$ \citep{krumholz2009ApJ...693..216K}. Hence, the very low depletion timescale or high star-formation efficiency for BCDs with W1$-$W2 $>$ 0.8 could be the result of runaway star-formation in dense, dust-rich and compact super star clusters, similar to what is seen in SBS 033$-$052, resulting in heated dust and low depletion timescales or high specific star-formation rates \citep{2004A&A...421..555H}.

\section{Summary and conclusion}
 \label{sec:summary}
In this paper, we present Arecibo H{\sc i} 21 cm observations of 11 MIR-bright BCDs, six of which resulted in detection.   We further combine these results with dIs/BCDs in the
literature and compare properties of MIR-bright dIs/BCDs with  dIs/BCDs that are not MIR-bright. We have the following conclusions from our study:
\begin{itemize}		
	\item MIR-bright dIs/BCDs follow the same H{\sc i} mass-stellar mass relation as other dIs/BCDs, showing H{\sc i} to stellar mass ratio decreasing with stellar mass and implying a similar gas-dominated  baryonic composition. We find that the $f_{\rm HI}$-$M_{\ast}$ relation shows stronger anti-correlation at  higher sSFR ($>$10$^{-8.5}$ yr$^{-1}$) implying that the current starburst episode in lower mass galaxies is mainly due to the higher gas fractions. In comparison to dI/BCDs of similar sSFR  in the literature, our sample of BCDs observed with the Arecibo telescope has higher stellar mass and hence lower H{\sc i} gas to stellar mass ratio, implying a lesser role of $f_{\rm HI}$ in the current starburst episode.
	
	\item We find a significant and strong anti-correlation of gas depletion timescales ($\tau_{\rm HI}$) with $M_{\ast}$ in high sSFR ($>$10$^{-8.5}$ yr$^{-1}$) dIs/BCDs and with sSFR in the sample of higher stellar mass ($>$10$^{7.5}$ M$_{\odot}$) objects. Strong $\tau_{\rm HI}$-sSFR  and $\tau_{\rm HI}$-$M_{\ast}$ anti-correlation indicates a more significant role of star-formation efficiency (SFE) in higher stellar mass and higher sSFR dIs/BCDs. This could imply a larger fraction of H{\sc i} gas being actively involved in star formation. MIR-bright dIs/BCDs have shorter gas depletion timescales or higher star formation efficiencies as compared to the other dIs/BCDs, especially  BCDs observed with the Arecibo have median depletion times scale of $\sim$0.3 Gyr, less by one order of magnitude compared to the median depletion time scale, $\sim$ 6.3 Gyr for other dIs/BCDs. 
	
	\item  For the  sources with W1$-$W2 $>$ 0.8, H{\sc i} depletion timescale are very low, $\sim$0.46 Gyr. Compared to other dIs/BCDs from the literature, BCDs in the Arecibo sample have the reddest W1$-$W2 colours. The majority of the BCDs in the Arecibo sample satisfy the criterion  W1$-$W2$>$ 0.8 implying the presence of warm dust. BCDs with redder W1$-$W2 colours are known to have high ionizing photon rates and compact starburst regions. However, the H{\sc i} gas fraction for the Arecibo sample is similar to the other dIs/BCDs of similar stellar mass in literature,  implying  no  change in the baryonic composition due to  ionizing photons. This could be due to the dense star-forming gas shielding the atomic H{\sc i} gas from excessive ionization and creating physical conditions favourable for faster H{\sc i} to H$_{2}$ conversion. Hence the high star formation efficiency or low depletion time scales are possibly due to the compact, dense and dust-rich star forming super clusters resulting in high sSFR.
	
\end{itemize} 
 Future H{\sc i} surveys with SKA pathfinder telescopes like Five hundred metre Aperture radio Telescope \citep[FAST; ][]{nan11, 2016RaSc...51.1060L, li18},  will probe  larger numbers of   galaxies with similarly high sSFR, and will provide definitive tests of  the results found here, many of which are marginal or just hinted at. With larger sky coverage and better H{\sc i} sensitivity than those of Arecibo \citep{2019SCPMA..6259506Z}, FAST can probe BCDs of lower stellar masses and higher sSFR. 
Such a sample will shed significant light on the characteristics of BCDs, particularly, their star formation history.   
\section*{Acknowledgements}
We thank an anonymous reviewer for useful comments which helped to improve the paper significantly.
This work is supported by the National Natural Science Foundation of China (NSFC) Grant No. 11988101, No. 12050410259, No. 11550110181, No. 11725313, No. 12041302, the International Partnership Program of Chinese Academy of Sciences grant No. 114A11KYSB20160008. YC was sponsored by the Chinese Academy of Sciences Visiting Fellowship for Researchers from Developing Countries, Grant No. 2013FFJB0009, by the FAST distinguished young researcher fellowship (19-FAST-02) from the Center for Astronomical Mega-Science, CAS, and Ministry of Science and Technology (MOST) grant no. QNJ2021061003L.  Y.P. acknowledge National Science Foundation of China (NSFC) Grant No. 12125301, 12192220, and 12192222, and the science research grants from the China Manned Space Project with NO. CMS-CSST-2021-A07. YZM acknowledges the support of the National Research Foundation with Grant No. 150580, No. 120385, and No. 120378. Portions of this research were carried out at the Jet Propulsion Laboratory, California Institute of Technology, under a contract with the National Aeronautics and Space Administration. 

We thank the staff of Arecibo Observatory (AO) which made these observations possible. During 2011-2018, the AO was  operated by SRI International under a cooperative agreement with the National Science Foundation (AST-1100968), and in alliance with Ana G. M\'endez-Universidad Metropolitana, and the Universities Space Research Association.  Currently, the AO is a facility of the National Science
Foundation operated under cooperative agreement (AST- 1744119) by the University of Central Florida in alliance with Universidad Ana G. Méndez (UAGM) and Yang Enterprises (YEI), Inc. YC thanks Priyadarshini Bangale for helping with the script to improve one of the figures.
This work has also used different Python packages e.g. NUMPY, ASTROPY, 
SCIPY and MATPLOTLIB. We thank numerous contributors for these packages. This research has made use of the SIMBAD database, operated at CDS, Strasbourg, France. This research has made use of the VizieR catalogue access tool, CDS, Strasbourg, France (DOI : 10.26093/cds/vizier). The original description  of the VizieR service was published in 2000, A\&AS 143, 23

This publication makes use of data products from the \textit{Wide-Field Infrared Survey Explorer}, which is a joint project of the University of California, Los Angeles, and the Jet Propulsion Laboratory, California Institute of Technology, funded by the National Aeronautics and Space Administration.  

This publication makes use of data products from the Two Micron All Sky Survey, which is a joint project of the University of Massachusetts and the Infrared Processing and Analysis Center/California Institute of Technology, funded by the National Aeronautics and Space Administration and the National Science Foundation.

This work also makes use of Sloan Digital Sky Survey (SDSS)-III. Funding for SDSS-III has been provided by the Alfred P. Sloan Foundation, the Participating Institutions, the National Science Foundation and the US Department of Energy Office of Science. The SDSS-III web site is http://www.sdss3.org/. SDSS-III is managed by the Astrophysical Research Consortium for the Participating Institutions of the SDSS-III Collaboration including the University of Arizona, the Brazilian Participation Group, Brookhaven National Laboratory, Carnegie Mellon University, University of Florida, the French Participation Group, the German Participation Group, Harvard University, the Instituto de Astrofisica de Canarias, the Michigan State/Notre Dame/JINA Participation Group, Johns Hopkins University, Lawrence Berkeley National Laboratory, Max Planck Institute for Astrophysics, Max Planck Institute for Extraterrestrial Physics, New Mexico State University, New York University, Ohio State University, Pennsylvania State University, University of Portsmouth, Princeton University, the Spanish Participation Group, University of Tokyo, University of Utah, Vanderbilt University, University of Virginia, University of Washington and Yale University.

This work also makes use of data products from the Panoramic Survey Telescope and Rapid Response System (Pan-STARRS). The Pan-STARRS1 Surveys were made possible through contributions by the Institute for Astronomy, the University of Hawaii, the Pan-STARRS Project Office, the Max-Planck Society and its participating
institutes, the Max Planck Institute for Astronomy, Heidelberg and the
Max Planck Institute for Extraterrestrial Physics, Garching, The Johns
Hopkins University, Durham University, the University of Edinburgh,
the Queen's University Belfast, the Harvard-Smithsonian Center for
Astrophysics, the Las Cumbres Observatory Global Telescope Network
Incorporated, the National Central University of Taiwan, the Space
Telescope Science Institute, and the National Aeronautics and Space
Administration under Grant No. NNX08AR22G issued through the Planetary
Science Division of the NASA Science Mission Directorate, the National
Science Foundation Grant No. AST-1238877, the University of Maryland,
Eotvos Lorand University (ELTE), and the Los Alamos National
Laboratory. The Pan-STARRS1 Surveys are archived at the Space Telescope
Science Institute (STScI) and can be accessed through MAST, the
Mikulski Archive for Space Telescopes. Additional support for the
Pan-STARRS1 public science archive is provided by the Gordon and Betty
Moore Foundation.

\section*{Data availability}
H{\sc i} 21-cm data from the Arecibo observations can be accessed using proposal ID A2715  at \url{https://www.naic.edu/datacatalog/}. Multi-band photometric data for UV (GALEX), optical (Pan-STARRS, SDSS), and infrared (2MASS, UKIDSS and AllWISE) are publicly available and can be accessed using VizieR catalogue service \url{https://vizier.cds.unistra.fr/viz-bin/VizieR}. The data from \cite{2012AJ....143..133H} (J/AJ/143/133) and \cite{2020AJ....160..271D} (J/AJ/160/271) are also publicly available and can be accessed using VizieR catalogue service. Stellar masses  and star formation rates  from MPA-JHU group are also publicly available at \url{http://wwwmpa.mpa-garching.mpg.de/SDSS/DR7/Data/stellarmass.html} and \url{http://wwwmpa.mpa-garching.mpg.de/SDSS/DR7/sfrs.html}. 

\appendix 
\section{}
	
\subsection{The location of H{\sc i} gas}
\label{hilocation}
\begin{figure}
	\includegraphics[scale=0.5]{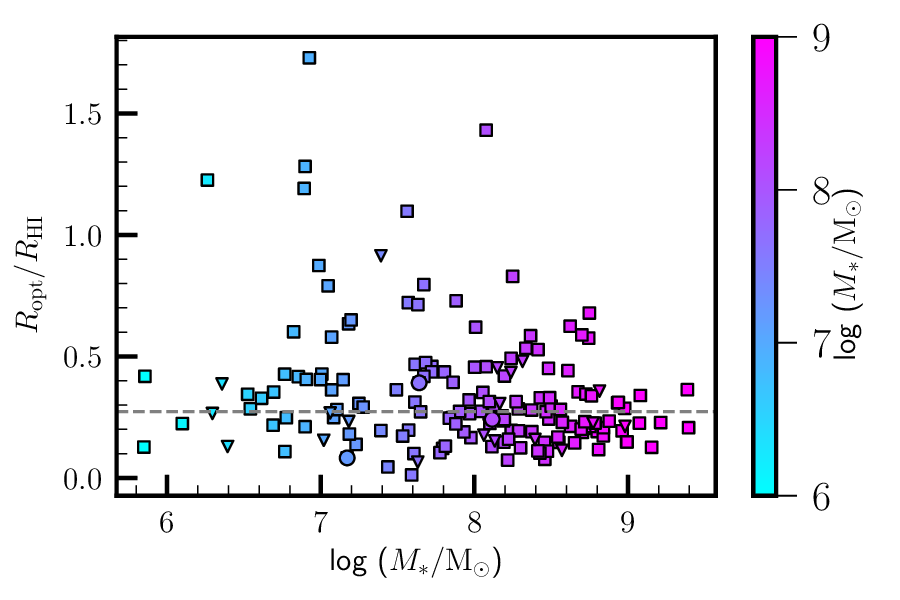}
	\includegraphics[scale=0.5]{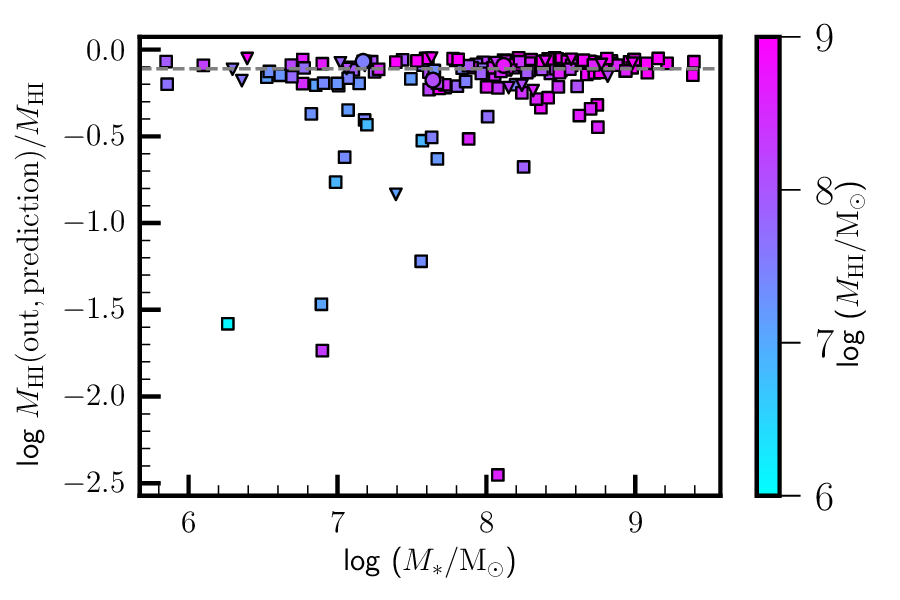}
	\caption{\textbf{Top panel:} $R_{\rm opt}$/$R_{\rm HI}$ vs $M_{\ast}$ . The symbols mean the same as in Fig.~\ref{wisecolor}. The horizontal grey dashed line shows the median value of $R_{\rm opt}$/$R_{\rm HI}$. \textbf{Bottom panel:} $M_{\rm HI}$(out, prediction)/$M_{\rm HI}$ vs $M_{\ast}$. The horizontal grey dashed line shows the median value of log($M_{\rm HI}$(out, prediction)/$M_{\rm HI}$).}
	\label{radiusratio}
\end{figure}

 Several previous studies found that H{\sc i} gas has a larger extent than the optical disk of galaxies \citep{wangjing2014MNRAS.441.2159W, wangjing2016MNRAS.460.2143W,wangjing2020ApJ...890...63W,hunt2020A&A...643A.180H}. For our study, it is important to check  whether the H{\sc i} measurement is coming from a similar H{\sc i} distribution relative to the optical disk.  Fig.~\ref{radiusratio} (upper panel) shows the variation of optical-to-H {\sc i} disk radius ratio versus stellar mass for our sample. H{\sc i} disk radius ($R_{\rm HI}$) is the semi-major axis of isophote where the  H{\sc i} gas column density is 1 M$_{\odot}$ pc$^{-2}$, and estimated it using the H{\sc i} mass-size relation from \cite{wangjing2016MNRAS.460.2143W}.  We used optical  major axis values   from the SIMBAD database to estimate the $R_{\rm opt}$ for 168 sources. We find that the optical-to-H{\sc i} radius ratio does not vary with stellar mass and has  a median value of $\sim$0.27. In Fig~\ref{radiusratio}  (lower panel), we follow \cite{wangjing2016MNRAS.460.2143W} and \cite{hunt2020A&A...643A.180H} to predict the H{\sc i} mass outside the optical radius, $M_{\rm HI }$ (out, prediction).  We find that the ratio of $M_{\rm HI }$ (out, prediction) to  $M_{\rm HI}$ does not systematically depend on stellar mass. The median logarithmic value of $M_{\rm HI}$(out, prediction)/$M_{\rm HI}$ is $\sim -$0.11, implying $\sim$78 \% of H{\sc i} gas is outside the stellar disk.

\clearpage
\onecolumn
\begin{deluxetable}{llcllcccc}
	\tablewidth{0pt}

	\tabletypesize{\scriptsize}
	\tablecaption{Sample of 44 MIR-bright dIs/BCDs. \label{mirBCDs}} 
	
	\startdata

		\hline
		(1)&(2)&(3)&(4)&(5)&(6)&(7)&(8)&(9)\\
		Source & Redshift & log $M \rm(HI)$ (Ref.) & telescope & log SFR (ref.)& log $M_{\ast}$ (ref.) & log sSFR & log $f_{\rm HI}$ & log $\tau_{\rm HI}$ \\
		&&[${\rm M}_{\odot}$]&&[${\rm M}_{\odot}$ yr$^{-1}$]&[${\rm M}_{\odot}$ ]& [yr$^{-1}$]&&[yr]\\ 
		\hline
0122+0743, UGC 993 		& 0.00975 	        & 9.47  (1)`  	& NRT 		& $-$1.17    (1)  & 6.39  (1) &$-$7.57 & 3.07 & 10.64 \\
W0231+2441    			& 0.0288 		& 8.30  (2)  		& Arecibo 	& $-$0.12    (2)  & 7.50  (2) & $-$7.62 & 0.80 & 8.42 \\
0335$-$052 			& 0.01349 		& 8.98  (3)  		& NRT 		& 0.03 (2)  & 7.62  (2) & $-$7.59 & 1.45 &9.03 \\
II Zw 40 			& 0.00267 	& 8.36  (4)	 	& Arecibo 	& $-$0.22   (3)  & 6.96  (2) & $-$7.18 & 1.61 & 8.79 \\
0749+568 			& 0.0183 		& 8.90  (3)  		& NRT   	& $-$0.45    (2)  & 8.10  (2) & $-$8.55 & 0.83 & 9.39 \\
W0801+2640 			& 0.026 		& 8.80  (2)  		& Arecibo 	& 0.36       (2)  & 7.97  (2) & $-$7.61 & 0.83 & 8.43 \\
W0830+0225 			& 0.0315 		& 9.10  (2)  		& Arecibo 	& 0.95       (2)  & 8.89  (2) & $-$7.94 & 0.21 & 8.15 \\
Mrk 108				& 0.00477 		& 8.20  (5)	  	& GMRT 		& $-$0.95    (2)  & 8.24  (2) & $-$9.19 & $-$0.11 & 9.08 \\
Mrk 22 				& 0.00539 		& 8.57  (6)	 	& GMRT 		& $-$1.33    (2)  & 8.13  (2) &$-$9.46 & 0.43 & 9.89 \\
W1016+3754 			& 0.0039 		& 7.50  (2)  		& Arecibo 	& $-$1.40    (2)  & 7.17  (2) & $-$8.57 & 0.32 & 8.90 \\
Mrk 140				& 0.00556 		& 8.77  (7)	  	& GBT 		& $-$0.95    (2)  & 8.57  (2) & $-$9.52 & 0.24 & 9.76 \\
SBS 1037+494 			& 0.00518 	& 8.28  (8)	  	& Effelsberg  	& $-$1.52 (2)  & 6.55  (2) & $-$8.07 & 1.77 & 9.84 \\
Mrk 724 			& 0.00406 		& 6.96  (4)	  	& Arecibo 	& $-$1.34    (2)  & 7.39  (2) & $-$8.73 & $-$0.30 & 8.43 \\
J1044+0353 			& 0.01315 		& 8.40  (9)	  	& GBT 		& $-$0.74 (2)  & 7.18  (2) & $-$7.93 & 1.27 & 9.19 \\
Haro 3 				& 0.00314 		& 8.73  (7) 	  	& GBT 		& $-$0.31    (4)  & 8.77  (2) & $-$9.08 & $-$0.07 & 9.01 \\
Mrk 1446			& 0.00988 		& 8.54  (8)   	& Effelsberg  	& $-$0.80    (2)  & 8.31  (2) & $-$9.11 & 0.26 &9.37 \\
Wa 22 				& 0.00472 		& 8.83  (4)	  	& Arecibo 	& $-$1.49    (2)  & 7.63  (2) &$-$9.12 & 1.31 & 10.42 \\
UM 439 				& 0.00368 		& 8.40  (4) 	& Arecibo 	& $-$1.16 (2)  & 7.15  (2) & $-$8.31 & 1.45 & 9.76 \\
Mrk 1450 			& 0.00311 		& 7.30  (8)	  	& Effelsberg  	& $-$1.46    (3)  & 7.31  (2) & $-$8.77 & 0.01 & 8.78 \\
Mrk 750 			& 0.00252 		& 7.42  (4)	  	& Arecibo	& $-$1.61    (3)  & 7.02  (2) & $-$8.63 & 0.57 & 9.20 \\
UM 461 				& 0.00350 & 8.04 (10)	 	& NRT 		& $-$1.48  (2)  & 7.06  (2) & $-$8.54 & 1.05& 9.59 \\
1152+579 			& 0.01726 		& 8.95  (3)  		& NRT 		& 0.09       (2)  & 8.08  (2) & $-$7.99 & 0.89 & 8.88 \\
J1201+0211 			& 0.00327 		& 7.18  (9)	  	& GBT 		& $-$1.89    (2)  & 6.36  (2) & $-$8.25 & 1.24 & 9.49 \\
J1202+5415 			& 0.01055 		& 7.79  (9)	  	& GBT 		& $-$1.32    (3)  & 6.29  (3) & $-$7.62 & 1.36 & 8.97 \\
Haro 06 			& 0.00673 		& 8.54  (4)	  	& Arecibo 	& $-$0.57    (2)  & 8.40  (2) & $-$8.97 & 0.26 & 9.23 \\
Mrk 49 				& 0.00510 		& 8.30  (4) 	  	& Arecibo 	& $-$0.65    (4)  & 8.01  (2) & $-$8.66 & 0.43 & 9.09 \\
Wa 53 				& 0.00865 		& 8.16  (4)	  	& Arecibo 	& $-$0.94    (2)  & 8.15  (2) & $-$9.09 & 0.09 & 9.18 \\
I Zw 36 			& 0.00094		& 7.82  (11)	 	& VLA 		& $-$1.81    (2) & 7.02  (2) & $-$8.83 & 0.80 & 9.63 \\
CG 184 				& 0.00313 	& 8.30  (4)	 	& Arecibo 	& $-$1.56    (3)  & 7.22  (2) & $-$8.78 & 1.13 & 9.91 \\
Wa 69 				& 0.0147 		& 8.78  (4)	  	& Arecibo 	& $-$0.54    (2)  & 8.17  (2) & $-$8.70 & 0.68 & 9.38 \\
Mrk 67 				& 0.0032 		& 7.41  (4)	  	& Arecibo 	& $-$1.63    (2)  & 7.46  (2) & $-$9.09 & $-$0.05 & 9.04 \\
Mrk 1369 			& 0.01194 		& 8.86  (4)	  	& Arecibo 	& $-$0.28    (2)  & 8.98  (2) & $-$9.27 & $-$0.08 & 9.19 \\
W1408+1759 			& 0.0237 	        & $<$8.20  (2)	 	& Arecibo 	& $-$0.01    (2)  & 7.69  (2) & $-$7.70 & 0.51 & 8.21 \\
W1423+2257 			& 0.0329 		& $<$8.70  (2)	 	& Arecibo 	& $-$0.14    (2)  & 8.12  (2) & $-$8.25 & 0.58 & 8.83 \\
SBS 1428+457 			& 0.00782 		& 9.01  (8)	 	& Effelsberg   	& $-$0.81    (3)  & 8.55  (2) & $-$9.37 & 0.47 & 9.84 \\
W1439+1702 			& 0.0301 		& $<$8.30 (2)	 	& Arecibo 	& $-$0.10    (2)  & 7.68  (2) & $-$7.78 & 0.62 & 8.39 \\
II Zw 070 			& 0.00401 		& 8.36  (4)	 	& Arecibo 	& $-$0.77    (2)  & 8.06  (2) & $-$8.83 & 0.28 & 9.11 \\
1629+205 			& 0.01719 		& 9.25  (3) 		& NRT   	& 0.62       (2)  & 9.30  (2) & $-$8.68 & $-$0.02 & 8.66 \\
Mrk 1499 			& 0.00897 		& 8.18  (8)	 	& Effelsberg   	& $-$0.74    (3)  & 8.81  (2) & $-$9.56 & -0.63 & 8.93 \\
2116+020, Mrk 513 		& 0.01818               & 9.58  (3) 		& NRT    	& $-$0.02    (2)  & 9.61  (2) & $-$9.63 & 0.04 & 9.67 \\
W2130+0830 			& 0.026 		& $<$8.10  (2)	 	& Arecibo 	& 0.12       (2)  & 7.74  (2) & $-$7.62 & 0.36 & 7.98 \\
W2212+2205 			& 0.0286 		& $<$8.50  (2)	 	& Arecibo 	& $-$0.16    (2)  & 8.07  (2) & $-$8.23 & 0.43 & 8.66 \\
W2238+1400 			& 0.0206 		& 8.10  (2)	 	& Arecibo 	& $-$0.26    (2)  & 7.64  (2) & $-$7.90 & 0.46 & 8.36 \\
W2326+0608 			& 0.01678 		& 7.80  (13)	 	& GMRT         	& $-$0.52    (3)  & 6.97  (2) & $-$7.49 & 0.83 & 8.31 \\

		\hline
	\enddata
		
	\begin{center}
		Notes: Different columns are as follows: (1) Source name, (2) redshift (3) logarithmic value of M(H{\sc i}) in units of $\rm M_{\odot}$, (4) telescope used, columns (5) to (9) are logarithmic values for star formation rate (SFR), stellar mass ($M_{\ast}$), specific star formation rate (sSFR), ratio of H{\sc i} mass to stellar mass($f_{\rm HI}$) and depletion time scale $\tau_{\rm HI}$.
	
	\begin{flushleft}
		References for H{\sc i} mass:(1) \cite{2007A&A...464..859P}, (2) This paper, (3) \cite{1999A&AS..139....1T}, (4) \cite{2002AJ....124..191S},  (5) \cite{2011ApJ...728..124R}, (6) \cite{2018MNRAS.473.4566P}, (7) \cite{1981ApJ...247..823T}, (8)\cite{2005A&A...434..887H},  (9) \cite{2016MNRAS.463.4268T}, (10)\cite{1999Ap.....42..149C} (11)\cite{2014A&A...566A..71L}, (12) \cite{2007A&A...462..919H}, (13) \cite{chandola2023MNRAS.523.3848C}. \\
		References for star formation rates:(1) MPA-JHU SDSS DR 7 catalog, (2) Estimated using FUV luminosity corrected for dust extinction using W4 values,  (3) Using predicted FUV luminosity from SED fitting and corrected for dust extinction using W4 values, (4)  Estimated using NUV luminosity corrected for dust extinction using W4 values. \\ 
		References for stellar mass:  (1) MPA-JHU SDSS DR 7 catalog, (2) this paper, (3) \cite{2016MNRAS.463.4268T}. \\
		All redshifts are from SIMBAD database except for those reported with H{\sc i} in this paper which are from Zhang Ludan et al. in preparation. H{\sc i} masses from the literature are re-estimated according to the cosmological luminosity distances  except for the sources from \cite{2014A&A...566A..71L} where the distances are estimated using tip of the red giant branch (TRGB) method.
		Abbreviations for different telescope names mean as follows: NRT: Nancay Radio Telescope, GBT: Green Bank Telescope, GMRT: Giant Metrewave Rado Telescope, VLA: Very Large Array \\
        
	\end{flushleft}
\end{center}
\end{deluxetable}
\clearpage
\newpage
\begin{deluxetable}{llcllcccc}	
	\tablewidth{0pt}
	\tabletypesize{\scriptsize}
	\tablecaption{Sample of other 185 dIs/BCDs detected with H{\sc i} from the literature. \label{otherBCDs} } 
	\startdata
\hline
Source & Redshift & log $M\rm(HI)$) (ref.)  & telescope & log SFR (ref.) & log $M_{\ast}$ (ref.) & log sSFR & log $f_{\rm HI}$ & log $\tau_{\rm HI}$ \\
	&&[${\rm M}_{\odot}$]&&[${\rm M}_{\odot}$ yr$^{-1}$]&[${\rm M}_{\odot}$] &[yr$^{-1}$]&&[yr]\\
\hline
AGC 748778 		& 0.00086 	& 6.67  (1) & VLA 		&$-$3.73 (1)    	& 6.00 (1)& $-$9.73 & 0.68 & 10.40 \\
0012$-$018, Mrk 546 	& 0.01311	& 8.68  (2)	& NRT 	 	& $-$0.60  (2)		& 8.80  (1)	& $-$9.40 & $-$0.12 & 9.28 \\
J0015+0104 		& 0.00685 	& 8.49  (3)	& Effelsberg    & $-$3.46  (3)	& 6.77  (1)	& $-$10.23 & 1.72 & 11.95 \\
UM 241 			& 0.01409	& 8.80  (4)	& Arecibo 	& $-$0.75  (2) & 8.70  (1)	& $-$9.45 & 0.09 & 9.55 \\
UM 38 			& 0.00463	& 8.40  (4)	& Arecibo 	& $-$1.47  (2)		& 7.79  (1)	& $-$9.27 & 0.61 & 9.88 \\
 UM 40 		 	&  0.00449 	&  8.80  (4)	& Arecibo 	&  $-$1.49  (2)	&  8.30  (1)	&  $-$9.79 &  0.50 &  10.29 \\
 J0031$-$0934,LEDA 989253&  0.01132 	&  8.65  (5)	& GBT 	        &  $-$1.83  (1)	&  7.72  (1)	&  $-$9.55 &  0.92 &  10.48 \\
 UM 51 			&  0.01412 	&  9.25  (4)	& Arecibo 	&  $-$1.16  (1)	&  8.74  (1)	&  $-$9.90 &  0.50 &  10.40 \\
UM 69 			& 0.00655 	& 9.07  (4)	& Arecibo 	& $-$1.08  (2)		& 8.84  (1)	& $-$9.92 & 0.23 & 10.15 \\
 UM 80 			&  0.01610 	&  9.25  (4)& Arecibo 	&  $-$1.13  (1)	&  8.48  (1)	&  $-$9.61 &  0.76 &  10.38 \\
0111+075, Mrk 564 	& 0.01848 	& 9.15  (2)	& NRT	 	& $-$0.65  (4) 		& 8.96  (1)	& $-$9.61 & 0.19 & 9.80 \\
 UM 92			&  0.02309 	&  8.64  (4)	& Arecibo 	&  $-$1.14  (1)	&  8.37  (1)	&  $-$9.50 &  0.27 &  9.77 \\
 UM 330			&  0.017 	&  8.73  (4)	& Arecibo 	&  $-$1.03  (1)	&  8.28  (1)	&  $-$9.31 &  0.45 &  9.75 \\
 UM 336			&  0.01924 &  8.94  (4) & Arecibo 	&  $-$1.29  (1)	&  8.22  (1)	&  $-$9.51 &  0.71 &  10.23 \\
J0133+1342 		& 0.00867 	& 7.83  (5)	& GBT 		& $-$1.77  (4) 	        & 6.94  (1)	& $-$8.71 & 0.89 & 9.59 \\
UM 345 			& 0.01891 	& 9.31  (4)	& Arecibo 	& $-$0.97  (2 )		& 8.49  (1)	& $-$9.46 & 0.83 & 10.29 \\
 AGC 110482 		&  0.00119 &  7.28  (1)	& VLA 		&  $-$3.22  (1) 	&  6.82  (1) 	&  $-$10.04 &  0.46 &  10.50 \\
 UM 372, UGC 1297 	&  0.00569 &  8.80  (4)& Arecibo 	&  $-$1.79  (4)	        &  8.37  (1)	&  $-$10.16 &  0.43 &  10.59 \\
 AGC 111977 		&  0.00069 &  6.85  (1)	& VLA 		&  $-$2.75  (5)          &  7.57  (2)	&  $-$10.32 &  $-$0.72 &  9.60 \\
UM 408 			& 0.01172 	& 8.87  (4)	& Arecibo 	& $-$1.08  (2)		& 8.15  (1)	& $-$9.23 & 0.72 & 9.95 \\
 UM 417 			&  0.00926 	&  8.21  (4)	& Arecibo 	&  $-$1.47  (2)          &  7.62  (1)	&  $-$9.08 &  0.60 &  9.68 \\
Mrk 370			& 0.00252 	& 8.63  (4)	& Arecibo 	& $-$1.32  (4)	        & 8.46  (1)	& $-$9.78 & 0.18 & 9.96 \\
Mrk 600 		& 0.00338 	& 8.56  (4)	& Arecibo 	& $-$1.51  (2)		& 7.88 	(1)     & $-$9.40 & 0.68 & 10.07 \\
 UGC 3672A 		&  0.00329 	&  8.65 (6)	& GMRT 	        &  $-$1.96  (1)         &  6.90  (1) 	&  $-$8.86 &  1.75 &  10.61 \\
 NGC 2366		&  0.00034 &  8.79 (7)	& VLA 		&  $-$1.57  (1)         &  8.41  (3)     &  $-$9.99 &  0.37 &  10.36 \\
 AGC 174585 		&  0.00119 &  6.90 (1) & VLA 		&  $-$2.91  (5)	       &  6.95  (2)    &  $-$9.86 &  $-$0.05 &  9.81 \\
 AGC 174605 		&  0.00117 	&  7.27 (1)	& VLA 		&  $-$3.07  (1)	&  6.34  (1)	&  $-$9.41 &  0.93 &  10.34 \\
HS 0822+3542 		& 0.00269 	& 7.93 (8)	& GMRT  	& $-$2.44 (2)   	& 5.85  (1)	& $-$8.29 & 2.08 & 10.37 \\
 J0843+4025 		&  0.00207 	&  6.68  (5)& GBT 		&  $-$2.88  (1)	       &  6.66  (1)	&  $-$9.73 &  0.02 &  9.75 \\
0846+3522 		& 0.00831 	& 7.48 (9)      & NRT 		& $-$1.89 (4) 	        & 7.18  (1) 	& $-$9.07 & 0.30 & 9.37 \\
 AGC 182595 		&  0.00133	&  7.00  (1)	& VLA 		&  $-$2.91  (1)	&  7.06  (1)	&  $-$9.97 &  $-$0.06 &  9.91 \\
0847+612, Mrk 99 	& 0.01282 	& 9.21  (2)	& NRT	 	& $-$0.57  (2)		& 8.99  (1)	& $-$9.57 & 0.22 & 9.78 \\
Mrk 16 			& 0.00777	& 8.94  (10)    & GBT 	        & $-$0.81  (2)		& 8.71  (1)	& $-$9.52 & 0.23 & 9.75 \\
 J0859+3923		&  0.0019 	&  7.10  (5) & GBT 		&  $-$3.19  (1)		&  6.92  (1)	&  $-$10.04 &  0.25 &  10.29 \\
CG 6 			& 0.00656 	& 8.42  (4)	& Arecibo 	& $-$1.02  (4)	        & 8.93  (1)	& $-$9.95 & $-$0.51 & 9.44 \\
J0903+0548 		& 0.0129 	& 8.34  (5)	& GBT 		& $-$1.42  (2)		& 8.05  (1)	& $-$9.47 & 0.29 & 9.76 \\
 J0908+0517 		&  0.00199	&  7.66  (5)	& GBT 		&  $-$3.25  (1)	&  5.86  (1)	&  $-$9.11 &  1.81 &  10.91 \\
 CG 10 			&  0.00634 &  8.10  (4)	& Arecibo 	&  $-$1.92  (1)	&  7.54  (1)	 &  $-$9.46 &  0.56 &  10.02 \\
 CG 13 			&  0.00627 	&  9.26  (4)& Arecibo	&  $-$1.21  (1)	       &  7.68  (1)	&  $-$8.90 &  1.57 &  10.47 \\
CG 14 			& 0.00631 	& 8.22  (4)	& Arecibo	& $-$1.23  (6)	        & 8.43  (1)	& $-$9.66 & $-$0.21 & 9.44 \\
 Mrk 104		&  0.00738 	&  8.73  (10)    & GBT 	        &  $-$0.73  (2)	       &  8.84  (1)	&  $-$9.57 &  $-$0.11 &  9.46 \\
Mrk 1416 		& 0.00779 	& 8.71  (11)	& Effelsberg    & $-$1.07  (2) 	        & 7.54  (1) 	& $-$8.61 & 1.17 & 9.78 \\
 J0921+3944 		&  0.01397	&  9.31  (5) & GBT 		&  $-$1.29  (1)	&  7.89  (1)	&  $-$9.18 &  1.42 &  10.60 \\
 I Zw 18 		&  0.00253 &  8.32  (7)	& VLA		&  $-$1.26  (2) 	        &  6.90  (1)	&  $-$8.15 &  1.42 &  9.58 \\
 0937+2949		&  0.00168 	&  7.40  (9)    & NRT		&  $-$2.46  (1)         &  6.91  (1)	&  $-$9.37 &  0.49 &  9.86 \\
 AGC 198691 		&  0.00172 	&  6.82  (12)    & Arecibo	&  $-$3.68  (1)		&  5.20  (1)	&  $-$8.88 &  1.63 &  10.51 \\
0940+4025 		& 0.01883 	& 8.50  (9)	& NRT 		& $-$0.96  (2)		& 7.38  (1)	& $-$8.34 & 1.13 & 9.46 \\
 0940+544 		&  0.00554	&  8.61  (2)	& GBT 		&  $-$1.85  (2)          &  7.59  (1)	&  $-$9.45 &  1.02 &  10.46 \\
 J0944+0936		&  0.00171	&  7.74  (5)	& GBT		&  $-$2.35  (1)	&  7.95  (2)	&  $-$10.30 &  $-$0.21 &  10.09 \\
SBS 0943+543 		& 0.0053 	& 7.07 (11)     & Effelsberg    & $-$1.58 (3)	        & 6.89  (4)	& $-$8.48 & 0.18 & 8.65 \\
SBS 0943+563B 		& 0.02 		& 9.57 (11)     & Effelsberg    & $-$0.74  (2)  	& 8.63  (1) 	& $-$9.37 & 0.94 & 10.31 \\
Mrk 407 		& 0.00529 	& 9.16 (10)     & GBT 	        & $-$1.20  (2)		& 8.46  (1)     & $-$9.66 & 0.71 & 10.36 \\
Mrk 1426 		& 0.00618 	& 8.25 (11)     & Effelsberg    & $-$1.41  (6)	        & 6.10  (4)	& $-$7.51 & 2.15 & 9.65 \\
 CG 34 			&  0.01730	&  8.43  (4)	& Arecibo 	&  $-$0.72  (2)	&  8.19  (1)	&  $-$8.91 &  0.24 &  9.15 \\
Haro 22 		& 0.00484 	& 8.04  (4)	& Arecibo 	& $-$1.75  (4)	        & 7.86  (4)	& $-$9.61 & 0.18 & 9.79 \\
Mrk 714 		& 0.00422 	& 7.93  (4)	& Arecibo 	& $-$1.66  (2)  	& 8.24  (1) 	& $-$9.90 & $-$0.31 & 9.59 \\
Haro 23, UGCA 201 	& 0.00457 	& 8.06  (4) 	& Arecibo 	& $-$1.08 (2) 		& 8.61 (4) 	& $-$9.69 & -0.55 & 9.14 \\
CG 55, KUG 1006+322 	& 0.00485 	& 8.22  (4) 	& Arecibo 	& $-$1.88 (2) 		& 7.98  (1)	& $-$9.85 & 0.24 & 10.09 \\
Wa 5 			& 0.00436 	& 7.39 (4)	& Arecibo 	& $-$1.81 (2) 	        & 7.41  (1)	& $-$9.22 & $-$0.02 & 9.20 \\
Wa 6 			& 0.01795 	& 9.30 (4) 	& Arecibo 	& $-$0.67 (2) 		& 8.75  (1)	& $-$9.42 & 0.54 & 9.97 \\
 Mrk 27			&  0.00711 	&  8.52 (10)     & GBT		&  $-$1.61 (1) 		&  8.19  (1)	&  $-$9.80 &  0.33 &  10.13 \\

 d1012+64 		&  0.00062	&  6.02  (13)	& GMRT 		&  $-$3.43  (1)	&  6.93  (1) &  $-$10.36 &  $-$0.91 &  9.45 \\

 Wa 8 			&  0.0036 	&  8.17  (4)	& Arecibo 	&  $-$1.39 (2)	&  8.10  (1)	&  $-$9.48 &  0.07 &  9.55 \\
 Mrk 32 			&  0.00283 &  7.96  (10)   & GBT 		&  $-$2.11 (2)	        &  7.44  (1)	&  $-$9.55 &  0.53 &  10.08 \\

 d1028+70 		&  $-$0.00036 	&  6.11  (13)	& GMRT		&  $-$3.44 (1) 	&  6.26 	(1)	&  $-$9.70 & $-$0.15 &  9.55 \\

 AGC 731457		&  0.00152	&  7.26  (1)	& VLA 		&  $-$2.49  (1)	&  7.15  (1)    &  $-$9.64 &  0.11 &  9.75 \\
 Wa 13 			&  0.00181 &  9.00  (4)	& Arecibo 	&  $-$1.37 (2)	        &  8.48  (1)&  $-$9.85 &  0.53 &  10.37\\
1033+4757 		& 0.0052 	& 8.04 (9)	& NRT		& $-$2.03  (2)		& 7.08  (4)	& $-$9.11 & 0.95 & 10.07 \\
1033+531 		& 0.00318 	& 7.64 (14)     & Effelsberg	& $-$1.89  (2)          & 7.57 (1) 	& $-$9.47 & 0.07 & 9.53 \\
 VV794, Spider 		&  0.00210 	&  9.13  (10)    & GBT 		&  $-$1.05  (2)         &  8.98  (4)	&  $-$10.03 &  0.15 &  10.18 \\

Mrk 1263 		& 0.00442 	& 8.65  (4)	& Arecibo 	& $-$1.44  (2)		& 6.77  (4)	& $-$8.21 & 1.89 & 10.10 \\
 CG 72 			&  0.00512	&  8.49  (15)	& NRT	 	&  $-$1.55  (2)		&  8.06  (1)	&  $-$9.60 &  0.43 &  10.04 \\
Mrk 156			& 0.00463 	& 8.64  (10)    & GBT 	        & $-$1.21  (2)		& 8.04  (1)	& $-$9.26 & 0.60 & 9.86 \\
 Mrk 157 		&  0.00464 	&  9.07  (10)    & GBT 	        &  $-$0.84  (2)	        &  8.55  (1)	&  $-$9.39 &  0.52 &  9.91 \\
 J1055+5111 		&  0.00457 	&  7.58  (5)	& GBT 		&  $-$2.17  (1)		&  7.04  (1)	&  $-$9.21 &  0.55 &  9.75 \\
Mrk 1271 	    	& 0.00339 	& 7.47  (4)	& Arecibo	& $-$1.18  (2)		& 8.28  (1)	& $-$9.46 & $-$0.80 & 8.65 \\
CG 75 			& 0.00425 	& 7.84  (4)	& Arecibo 	& $-$1.92  (2)		& 7.93  (1)	& $-$9.85 & $-$0.09 & 9.76 \\
CG 76 			& 0.00552 	& 7.71  (4)	& Arecibo 	& $-$1.91  (2)		& 7.97  (1)	& $-$9.88 & $-$0.26 & 9.62 \\
SBS 1054+504 		& 0.00459 	& 7.51 (11)     & Effelsberg    & $-$1.75  (2)	        & 8.20  (1) 	& $-$9.94 & $-$0.69 & 9.25 \\
 1059+3934 		&  0.01098 	&  8.71 (9)	& NRT		&  $-$1.42  (1)		&  7.90  (1)	&  $-$9.32 &  0.80 &  10.12 \\
Haro 4 			& 0.00216 	& 7.55 (16)     & VLA 	        & $-$1.73  (2)		& 7.01  (1)	& $-$8.73 & 0.55 & 9.28 \\
SBS 1114+587 		& 0.00561 	& 7.86 (11)     & Effelsberg    & $-$1.61  (2)		& 7.97  (4)	& $-$9.58 & -0.11 & 9.47 \\
 J1119+0935 		&  0.00331 	&  7.97  (5)	& GBT 		&  $-$2.18  (1)		&  7.25  (1)	&  $-$9.43 &  0.72 &  10.15 \\
 SBS 1116+517 		&  0.00446 	&  8.05 (11)     & Effelsberg    &  $-$1.71  (1)		&  7.23  (1)	&  $-$8.94 &  0.82 &  9.76 \\
 J1121+5720 		&  0.00358 	&  7.37  (5)	& GBT 		&  $-$2.81  (1)	        &  7.05  (1) 	&  $-$9.86 &  0.33 &  10.19 \\
 J1121+0324 		&  0.00383 	&  8.18  (5)	& GBT 		&  $-$2.43 (1)		&  6.69  (1)	&  $-$9.12 &  1.49 &  10.61 \\

CG 103 			& 0.00541 	& 8.66  (4)	& Arecibo	& $-$1.05  (2)		& 8.49  (1)	& $-$9.53 & 0.17 & 9.71 \\
 1123+644 		&  0.00330 	&  8.55  (2)	& GBT 		&  $-$1.52 (2) 	        &  8.29  (1)	&  $-$9.81 &  0.26 &  10.07 \\
 J1127+6536		&  0.00411 	&  7.06  (5)	& GBT 		&  $-$2.83  (1)		&  7.11  (1)	&  $-$9.94 &  $-$0.05 &  9.89 \\
 J1128+5714 		&  0.00556 	&  7.94  (5)	& GBT 		&  $-$2.25  (1)		&  7.15  (1)	&  $-$9.40 &  0.80 &  10.20 \\
Mrk 424 		& 0.00647 	& 8.62  (4)	& Arecibo 	& $-$1.00  (6)	        & 8.72  (1)	& $-$9.72 & $-$0.10 & 9.62 \\
 1128+573 		&  0.00577 	&  7.94 (14)     & Effelsberg    &  $-$1.91 (2) 	        &  7.11 (1) 	&  $-$9.01 &  0.83 &  9.85 \\
 Mrk 178 		&  0.00078 	&  7.11 (17)	& GBT 		&  $-$2.39 (1) 		&  6.53  (1) 	&  $-$8.92 &  0.59 &  9.51 \\
 Wa 23 			&  0.00812 	&  8.86  (4)	& Arecibo 	&  $-$1.00 (2) 		&  8.68  (1)	&  $-$9.67 &  0.19 &  9.86 \\
SBS 1137+589 		& 0.00647 	& 8.13 (11)     & Effelsberg    & $-$1.94  (6)		& 7.61  (1)	& $-$9.55 & 0.52 & 10.07 \\
 Mrk 426 		&  0.00518 	&  8.84  (4)	& Arecibo 	&  $-$1.09  (2) 		&  8.54  (1)	&  $-$9.63 &  0.29 &  9.93 \\
 Wa 25 			&  0.00614 	&  8.60  (4)	& Arecibo 	&  $-$1.43  (4) 		&  7.28  (1)	&  $-$8.70 &  1.32 &  10.02 \\
Wa 27 			& 0.00608 	& 8.84  (4)	& Arecibo	& $-$1.22  (2)		& 8.43 (1)	& $-$9.64 & 0.41 & 10.06 \\
Wa 28 			& 0.00603 	& 8.98  (4)	& Arecibo 	& $-$1.41 (2)		& 7.78 (1)	& $-$9.19 & 1.21 & 10.39 \\
Mrk 747 		& 0.00250 	& 7.49  (4)	& Arecibo 	& $-$1.78  (2)		& 7.86 (1)	& $-$9.65 & $-$0.37 & 9.28 \\
CG 123 			& 0.00616 	& 8.37  (4)	& Arecibo 	& $-$1.05  (2)		& 8.57  (1)	& $-$9.62 & $-$0.21 & 9.42 \\
NGC 3870, Mrk 186 	& 0.00241 	& 8.17  (8)     & GBT 		& $-$1.23  (2) 		& 8.47  (1)	& $-$9.70 & $-$0.30 & 9.40 \\
 SBS 1144+591 		&  0.00944 	&  8.98 (11)     & Effelsberg 	&  $-$1.28  (1)		&  8.41 (1) 	&  $-$9.70 &  0.56 &  10.26 \\
1145+601 		& 0.00420 	& 8.06 (14)     & Effelsberg 	& $-$1.49  (2)		& 8.18  (1)	& $-$9.67 & $-$0.12 & 9.55 \\
J1148+5400 		& 0.00866 	& 8.45  (5)	& GBT 		& $-$1.50  (3)		& 7.70  (1)	& $-$9.20 & 0.75 & 9.95 \\
Wa 34 			& 0.0112 	& 7.88  (4)	& Arecibo 	& $-$1.22  (2)		& 8.25 (1)	& $-$9.46 & $-$0.37 & 9.10 \\
Mrk 1460 		& 0.00265 	& 6.80  (11)    & Effelsberg    & $-$2.41  (2)	        & 7.20  (1)	& $-$9.61 & $-$0.39 & 9.22 \\
Mrk 641 		& 0.00736 	& 8.07  (4)	& Arecibo 	& $-$1.20  (2)		& 8.37  (1)	& $-$9.57 & $-$0.30 & 9.27 \\
CG 130 			& 0.01091 	& 8.57  (4)	& Arecibo 	& $-$1.40  (2)		& 8.75  (1)	& $-$10.15 & $-$0.18 & 9.97 \\
 CG 142 			&  0.00206 	&  7.70  (4)	& Arecibo 	&  $-$2.09  (2) 		&  8.01  (1) 	&  $-$10.10 &  $-$0.31 &  9.78 \\
Mrk 756 		& 0.00495 	& 8.79  (4)	& Arecibo 	& $-$1.06  (2)		& 8.19 (1)	& $-$9.25 & 0.59 & 9.84 \\
CG 150 			& 0.00204 	& 7.50  (4)	& Arecibo 	& $-$2.01  (2)		& 7.84  (1)	& $-$9.84 & $-$0.34 & 9.51 \\
1203+592 		& 0.01092 	& 9.33  (2)	& NRT 		& $-$0.64  (2)		& 8.93  (1)	& $-$9.57 & 0.40 & 9.97 \\
SBS 1205+557 		& 0.0059 	& 7.52 (11) 	& Effelsberg	& $-$1.64 (2) 		& 7.63  (1)	& $-$9.27 & $-$0.11 & 9.16 \\
 CG 159 			&  0.02195 	&  8.54  (4)	& Arecibo 	&  $-$0.95 (2)		&  8.31  (1)	&  $-$9.26 &  0.23 &  9.49 \\
1208+531 		& 0.00303 	& 7.72 (14) 	& Effelsberg 	& $-$2.42 (2) 		& 7.11  (1)	& $-$9.53 & 0.60 & 10.13 \\
 NGC 4163 		&  0.00055 	&  7.18  (7) 	& VLA 		&  $-$2.65 (1)		&  7.49  (1)	&  $-$10.14 &  $-$0.31 &  9.83 \\
UM 483 			& 0.0078 	& 8.65  (4)	& Arecibo 	& $-$1.19 (2)		& 8.12  (1)	& $-$9.30 & 0.54 & 9.84 \\
SBS 1211+540 		& 0.00304 	& 7.30 (11)     & Effelsberg 	& $-$2.37 (2) 		& 6.54  (1)	& $-$8.91 & 0.76 & 9.67 \\
1212+505 		& 0.0134 	& 8.48 (14)     & Effelsberg 	& $-$1.37 (2) 		& 7.97  (1)	& $-$9.34 & 0.51 & 9.85 \\
 J1214+0940 		&  0.00566 	&  7.22 (5)	& GBT 		&  $-$2.37  (1)		&  7.67  (1)	&  $-$10.04 &  $-$0.46 &  9.58 \\
CG 165 			& 0.02837 	& 9.00 (4)	& Arecibo 	& $-$0.69  (2)		& 7.31  (1)	& $-$8.01 & 1.69 & 9.69 \\
 J1215+5223 		&  0.00051 	&  7.24 (5)& GBT 		&  $-$3.11  (1)		&  6.62  (1)	&  $-$9.72 &  0.63 &  10.35 \\

 Haro 28, NGC 4218 	&  0.00241 	&  8.15 (10)     & GBT           &  $-$1.41 (2) 		&  8.50  (1)     &  $-$9.91 &  $-$0.35 &  9.56 \\
1213+597 		& 0.01468 	& 8.96 (2)	& NRT	 	& $-$0.34  (2)		& 9.39  (1)	& $-$9.73 & $-$0.43 & 9.30 \\
 VV 432 &  $-$0.00056 	&  8.60 (10)     & Arecibo	&  $-$1.69  (1) 		&  8.08  (1)	&  $-$9.77 &  0.52 &  10.30 \\
UM 491 			& 0.00669 	& 8.26 (4)	& Arecibo 	& $-$1.13  (2)		& 8.10  (1)	& $-$9.23 & 0.16 & 9.39 \\
Wa 52 			& 0.00861 	& 9.09 (4)	& Arecibo 	& $-$1.10  (2)		& 8.22  (1)	& $-$9.32 & 0.87 & 10.19 \\
Mrk 1323 		& 0.00627 	& 8.28 (4)	& Arecibo 	& $-$1.34  (2)		& 8.56  (1)	& $-$9.90 & $-$0.28 & 9.62 \\
 AGC 749237 		&  0.00124 	&  7.76 (1)	& VLA 		&  $-$2.34  (7)		&  6.69  (1) 	&  $-$9.03 &  1.07 &  10.10 \\
 NGC 4449 		&  0.00068 	&  9.48 (7)	& VLA 		&  $-$0.50 (2) 		&  8.70  (1) 	&  $-$9.19 &  0.78 &  9.98 \\
 1227+563 		&  0.015 	&  8.70 (14)     & Effelsberg	&  $-$1.58 (1)	        &  8.13  (1) 	&  $-$9.71 &  0.57 &  10.28 \\
 CG 187 			&  0.01435 	&  9.21 (4)	& Arecibo 	&  $-$1.22  (1)		&  8.73  (1)	&  $-$9.94 &  0.49 &  10.43 \\
 CG 189 			&  0.03105 	&  9.38 (4)	& Arecibo 	&  $-$0.57  (2)		&  8.23  (1)	&  $-$8.80 &  1.15 &  9.95 \\
 UM 513 			&  0.01248 	&  8.54  (4)	& Arecibo 	&  $-$1.46  (1)		&  8.70  (1)	&  $-$10.15 &  $-$0.16 &  9.99 \\
1241+549 		& 0.01581 	& 9.28  (2)	& NRT 		& $-$1.32  (4) 		& 8.37  (1) 	& $-$9.69 & 0.91 & 10.60 \\
 Haro 33 		&  0.00316 	&  8.53 (4)	& Arecibo 	&  $-$1.46  (2)		&  8.10  (1)	&  $-$9.56 &  0.43 &  9.99 \\
Haro 9			& 0.00358 	& 8.93 (10)     & Arecibo 	& $-$0.47 (2)		& 8.81 (1)	& $-$9.29 & 0.12 & 9.41 \\
Mrk 224 		& 0.003 	& 7.35  (15)	& NRT 		& $-$1.74 (2)		& 7.56  (1)	& $-$9.30 & $-$0.21 & 9.09 \\
1248+518 		& 0.0112 	& 8.65 (14)	& Effelsberg 	& $-$0.99  (2)		& 8.49  (1)	& $-$9.48 & 0.16 & 9.64 \\
 Mrk 1338 		&  0.00362 	&  7.73  (4)	& Arecibo 	&  $-$2.26 (1)		&  8.29 (1) 	&  $-$10.55 & $-$0.56 &  9.99 \\
 VV 558 			&  0.00071 	&  6.85  (10)    & GBT 		&  $-$3.24  (1)		&  6.66  (1)	&  $-$9.90 &  0.19 &  10.09 \\
UM 538 			& 0.00313 	& 6.88  (4)	& Arecibo 	& $-$2.48  (2)		& 6.99  (1)	& $-$9.47 & $-$0.10 & 9.36 \\
1318+520 		& 0.01582 	& 9.30  (2)	& NRT	 	& $-$0.47  (2)		& 9.21 (1)	& $-$9.69 & 0.09 & 9.78 \\
1323+483 		& 0.01643 	& 8.46  (2)	& NRT 	 	& $-$0.84  (2)		& 8.06 (1)	& $-$8.90 & 0.40 & 9.30 \\
 J1328+6341 		&  0.00604 	&  7.49  (2)	& GBT 		&  $-$2.52  (1)		&  7.28 (1)	&  $-$9.80 &  0.21 &  10.01 \\
 CPG 384 		&  0.00080 	&  7.46 (10)	& GBT 		&  $-$2.30  (2) 		&  7.07  (1)	&  $-$9.37 &  0.38 &  9.75 \\
Haro 38 		& 0.00286 	& 8.01 (4)	& Arecibo 	& $-$1.83  (2) 		& 7.88  (1) 	& $-$9.71 & 0.13 & 9.85 \\
J1335+4910 		& 0.00211 	& 7.51 (5)	& GBT 		& $-$2.86  (4) 		& 6.86  (1)	& $-$9.71 & 0.66 & 10.37 \\
1341+594 		& 0.01025 	& 8.49 (2)	& NRT	 	& $-$1.33  (2)		& 7.72 (4)	& $-$9.06 & 0.77 & 9.83 \\
 Mrk 277 		&  0.00575 	&  9.07 (10) 	& GBT		&  $-$1.70 (4) 		&  7.61  (1)	&  $-$9.31 &  1.46 &  10.77 \\
UM 618 			& 0.01502	& 8.34 (4)	& Arecibo	& $-$1.17  (2)		& 8.08 (1)	& $-$9.25 & 0.26 & 9.51 \\
 Wa 81 			&  0.00650 	&  7.94 (4)& Arecibo 	&  $-$1.71  (2)		&  7.64  (1)	&  $-$9.35 &  0.30 &  9.65 \\
 1358+554E 		&  0.01382 	&  9.08 (2)	& NRT	 	&  $-$0.76  (1)		&  8.09  (1) 	&  $-$8.85 &  0.99 &  9.84 \\
1359+504 		& 0.00590 	& 7.94 (14) 	& Effelsberg	& $-$1.53  (2) 		& 8.27  (1) 	& $-$9.80 & $-$0.33 & 9.47 \\
 SBS 1400+461 		&  0.00704 	&  8.57 (11) 	& Effelsberg 	&  $-$0.92  (2) 		&  8.88 (1) 	&  $-$9.80 &  $-$0.31 &  9.49 \\
 SBS 1401+490 		&  0.00240 	&  7.24 (11)     & Effelsberg    &  $-$2.78  (1)   	&  7.00 (1) 	&  $-$9.78 &  0.24 &  10.02 \\
J1404+5114 		& 0.0059 	& 8.18 (5) 	& GBT 		& $-$1.91  (3) 		& 7.39 (1) 	& $-$9.30 & 0.79 & 10.09 \\
1413+495 		& 0.01284 	& 8.94 (2)	& NRT	 	& $-$1.28  (2)		& 8.20 (4) 	& $-$9.48 & 0.74 & 10.22 \\
1422+573 		& 0.01057 	& 9.17 (2)	& GBT   	& $-$0.74 (6)		& 8.99 (1) 	& $-$9.74 & 0.17 & 9.91 \\
 1430+596 		&  0.00604 	&  8.40  (2)	& NRT		&  $-$1.57 (2)		&  8.24 (1) 	&  $-$9.81 &  0.16 &  9.98 \\
SBS 1430+526		& 0.0109 	& 9.09 (11) 	& Effelsberg 	& $-$0.74 (6)		& 8.76 (1) 	& $-$9.50 & 0.32 & 9.83 \\
Mrk 1384 		& 0.00762 	& 9.02  (4)	& Arecibo 	& $-$1.66  (2)		& 7.81 (1) 	& $-$9.47 & 1.21 & 10.68 \\
Haro 43 		& 0.00637	& 8.94  (4)	& Arecibo 	& $-$1.24  (2)		& 7.19 (1) 	& $-$8.43 & 1.76 & 10.18 \\

 1435+516 		&  0.00783 	&  8.68 (14)     & Effelsberg    &  $-$1.42  (1) 		&  8.62 (1)	&  $-$10.14 &  0.05 &  10.10 \\
Mrk 475 		& 0.00186 	& 6.56  (18)	& GMRT 		& $-$2.40  (2) 		& 6.71 (1) 	& $-$9.10 & $-$0.14 & 8.96 \\
Mrk 826 		& 0.00239 	& 7.32 (11) 	& Effelsberg 	& $-$2.21  (2) 		& 7.65 (1) 	& $-$9.87 & $-$0.33 & 9.53 \\
II Zw 71 		& 0.00418 	& 8.87  (4)	& Arecibo	& $-$1.30  (2) 		& 8.65 (1) 	& $-$9.95 & 0.22 & 10.17 \\
SBS 1453+526 		& 0.01088 	& 8.57 (11) 	& Effelsberg	& $-$1.05 (2) 		& 8.34 (1) 	& $-$9.39 & 0.23 & 9.62 \\
1455+443 		& 0.01182 	& 9.08  (2) 	& GBT   	& $-$0.75  (2)		& 8.79 (1) 	& $-$9.53 & 0.30 & 9.83 \\
1506+553 		& 0.01118 	& 9.48  (2) 	& NRT 		& $-$0.59 (2) 		& 9.39 (1) 	& $-$9.99 & 0.09 & 10.07 \\
1509+527 		& 0.01183	& 8.61  (2)	& NRT 		& $-$0.83  (2)		& 8.94 (1) 	& $-$9.77 & $-$0.33 & 9.44 \\

 1510+571 		&  0.00215 	&  7.17 (14)  	& Effelsberg 	&  $-$2.89 (1) 		&  7.07 (1)	&  $-$9.96 &  0.10 &  10.05 \\
1519+496 		& 0.01511 	& 8.99  (2)	& NRT 		& $-$0.61  (2)		& 9.08 (1) 	& $-$9.69 & $-$0.09 & 9.60 \\
Mrk 850 		& 0.00734 	& 8.17  (4)	& Arecibo 	& $-$1.55 (4) 		& 7.67 (1) 	& $-$9.22 & 0.50 & 9.72 \\

 1540+576A &  0.0121 &  8.41 (14) &   Effelsberg &  $-$1.41 (2) &  8.48 (1) &  $-$9.89 &  $-$0.07 &  9.80 \\
1551+601A 		& 0.01040 	& 8.70  (2)	& NRT 		& $-$1.10 (2)		& 8.00 (1) 	& $-$9.10 & 0.70 & 9.80 \\

 1704+4332 		&  0.00695 	&  7.70 (9)	& NRT 		&  $-$1.80 (2) 		&  6.78 (1) 	&  $-$8.58 &  0.93 &  9.50 \\
1707+565 		& 0.01114 	& 8.63  (2)	& NRT 		& $-$0.77  (2)		& 8.72 (1) 	& $-$9.49 & $-$0.09 & 9.40 \\
1714+602 		& 0.02013 	& 9.27  (2)	& NRT		& $-$0.66  (2)		& 9.08 (1) 	& $-$9.74 & 0.19 & 9.93 \\

 J2053+0039 		&  0.01328 	&  9.09  (3)	& Effelsberg	&  $-$2.26  (1)		&  7.53 (1) 	&  $-$9.79 &  1.56 &  11.35 \\

 J2150+0033 		&  0.01508 	&  9.05  (3)	& Effelsberg	&  $-$2.08  (1)		&  7.39 (1) 	&  $-$9.47 &  1.65 &  11.13 \\
2246+315 		& 0.01285 	& 9.23  (2)	& NRT 		& $-$0.41  (2)		& 8.81 (1) 	& $-$9.22 & 0.42 & 9.64 \\

NGC 7468 		& 0.00695 	& 9.42  (10) 	& Arecibo 	& $-$0.45  (4) 		& 9.15 (1) 	& $-$9.61 & 0.27 & 9.87 \\

 Mrk 324 &  0.00538 	&  8.42  (4)	& Arecibo 	&  $-$1.28 (1)		&  8.22 (1) 	&  $-$9.50 &  0.20 &  9.70 \\

 \hline
	\enddata
\begin{center}
	Notes: Different columns are as follows: (1) Source name, (2) redshift, (3) logarithmic values of $M$(H{\sc i}) in units of $\rm M_{\odot}$, (4) telescope used, columns (5) to (9) are logarithmic values for star formation rate (SFR), stellar mass ($M_{\ast}$), specific star formation rate (sSFR), ratio of H{\sc i} mass to stellar mass($f_{\rm HI}$) and depletion time scale $\tau_{\rm HI}$.

\begin{flushleft}
	References for H{\sc i} mass: (1) \cite{2016ApJ...832...85T} , (2) \cite{1999A&AS..139....1T}, (3)\cite{2013A&A...558A..18F}, (4) \cite{2002AJ....124..191S}, (5)\cite{2016MNRAS.463.4268T},  (6)\cite{2017MNRAS.465.2342C},  (7)\cite{2014A&A...566A..71L},  
	(8)\cite{2006MNRAS.371.1849C},  (9)\cite{2007A&A...464..859P}, (10) \cite{1981ApJ...247..823T},  (11)\cite{2005A&A...434..887H}, (12) \cite{2016ApJ...822..108H},  (13)\cite{2012MNRAS.426..665R},   (14)\cite{2007A&A...462..919H},  (15)\cite{1999Ap.....42..149C},  (16)\cite{2004AJ....127..264B},  (17)\cite{2017AJ....153..132A},   (18)  \cite{2020MNRAS.498.4745J}\\

	References for star formation rates: (1) Estimated using FUV luminosity without any correction,
	(2) Estimated using FUV  luminosity corrected for dust extinction using W4 values, (3) MPA-JHU SDSS DR 7 catalog, (4) Using predicted FUV luminosity from SED fitting and corrected for dust extinction using W4 values, (5)\cite{2016ApJ...832...85T},  (6)  Estimated using NUV luminosity corrected for dust extinction using W4 values, (7) Estimated using NUV luminosity without any correction for dust extinction.\\ 
	
 References for stellar mass: (1) This paper,  (2)\cite{2016ApJ...832...85T}, (3)  \cite{2014A&A...566A..71L}, (4) MPA-JHU SDSS DR 7 catalog.\\
	All redshifts are from the SIMBAD database. 
   In our analysis, for the sources in \protect\cite{2012MNRAS.426..665R}, \protect\cite{2014A&A...566A..71L}, \protect\cite{2016ApJ...832...85T} and \protect\cite{2017AJ....153..132A}, we use  tip of the red giant branch (TRGB) distances used for estimating the H{\sc i} mass in these papers. For VV 432, we use the Tully-Fisher distance from the extragalactic distance database  \protect\citep{tully2009AJ....138..323T}. For VV 558, J1215+5223 and CPG 384, we use the TRGB distances from \protect\cite{mccall2012A&A...540A..49M}, \protect\cite{sabbi2018ApJS..235...23S} and  \protect\cite{tully2013AJ....146...86T} respectively.
  For rest of the sources, H{\sc i} masses from the literature are re-estimated according to  cosmological luminosity distances.
	Abbreviations for different telescope names mean as follows: NRT: Nancay Radio Telescope, GBT: Green Bank Telescope, GMRT: Giant Metrewave Rado Telescope, VLA: Very Large Array \\  
\end{flushleft}
\end{center}
\end{deluxetable}

\end{document}